\documentclass[%
 aip,
 jmp,%
 amsmath,amssymb,
preprint,%
]{revtex4-2}

\usepackage{graphicx}
\usepackage{dcolumn}
\usepackage{bm}
\DeclareMathSymbol{*}{\mathbin}{symbols}{"01}
\usepackage{float}
\usepackage{xcolor}
\usepackage{multirow}
\usepackage{longtable}
\usepackage{hyperref}

\usepackage{xr}
\begin{document}


\title[Benchmarking SOPPA-based methods]{Benchmarking SOPPA-based methods for the calculation of static and dynamic polarizabilities}

\author{Joep van den Brink}
\affiliation{ 
Department of Chemistry, University of Copenhagen, Universitetsparken 5, DK-2100 Copenhagen Ø
}
\author{Michael Hilding Estes}%
\affiliation{ 
Department of Chemistry, University of Copenhagen, Universitetsparken 5, DK-2100 Copenhagen Ø
}

\author{Stephan P. A. Sauer}
\email{sauer@chem.ku.dk}
\affiliation{ 
Department of Chemistry, University of Copenhagen, Universitetsparken 5, DK-2100 Copenhagen Ø
}

\date{\today}

\begin{abstract}
Static and frequency‑dependent polarizabilities were computed for 41 molecules using RPA, RPA(D), HRPA, HRPA(D), SOPPA, SOPPA(CC2), and SOPPA(CCSD) with the aug‑cc‑pVTZ basis set and benchmarked against CCSD reference values and available experimental data. The analysis reveals a pronounced distinction between the performance of these methods for aromatic versus non‑aromatic molecules. Across all frequencies, HRPA consistently yields substantially larger deviations from CCSD than the other approaches, whereas HRPA(D) and SOPPA(CCSD) provide the most accurate results overall. For static polarizabilities, HRPA(D) performs best for non‑aromatic systems, followed by SOPPA(CCSD) and RPA(D), while SOPPA(CCSD) is most accurate for aromatic molecules. In the frequency‑dependent regime, HRPA(D) remains the most accurate method for non‑aromatic molecules, although RPA(D) shows greater consistency. For aromatic molecules, SOPPA(CCSD) performs best at low frequencies, with RPA offering intermediate accuracy but higher consistency than most other methods; at higher frequencies, RPA becomes the most accurate approach, followed by RPA(D), while SOPPA(CCSD) deteriorates. These trends highlight the importance of doubles corrections in RPA(D) and HRPA(D), which achieve accuracy comparable to or better than SOPPA(CCSD) at lower computational cost. The strong performance of RPA for aromatic molecules is attributed to its characteristic overestimation of the lowest electronic excitation energy. Comparison with experimental data confirms SOPPA(CCSD) as the most reliable method for static polarizabilities, while RPA and HRPA(D) provide the best agreement for frequency‑dependent polarizabilities of aromatic systems.
\end{abstract}

\keywords{Polarizability, SOPPA, HRPA(D), RPA(D), aromatic molecules}
\maketitle

\section{\label{Intro}Introduction}
Studying the properties of molecules in an external electric field is of great importance to chemical and physical applications such as spectroscopy. \cite{ImpPolSpec,ImpPolIR,Polbook}
A significant property of the molecule is its dipole polarizability, and it is used in spectroscopic methods such as Raman. \cite{Ramanpol,RamanTheory,Svendsenraman,Nørbyraman} Experimental measurements of polarizabilities are being made, \cite{Heitzexppol,Furanexp2,Furanexp3,Exp589} however, calculating the polarizability is beneficial since experimental data are not always readily available. 
Methods based on Coupled Cluster theory have proven to be very accurate,\cite{CCSD, CClinear, CCpol, CCbench, spas229} but they tend to be computationally expensive. 
Therefore, methods of lower computational cost while retaining a similar accuracy are valuable.
Over time, several methods have been developed to achieve this, including empirical methods,\cite{newempmeth} time-dependent density functional theory,\cite{CasidaDFT,CalaminiciDFT,PeachDFT,DreuwDFT} perturbation theory methods,\cite{Oddershede, ADC, 1SOPPA, CC2, MP2der, QEDMP2, 4SOPPACCSD, MP2method} etc.

The simplest method of approximation is time-dependent Hartree-Fock theory (TD-HF), also known as the Random-Phase Approximation (RPA).\cite{RPA} 
Using a perturbation expansion, RPA is correct through first order in the electronic repulsion. 
At this level of approximation, only single excitation- and de-excitation operators are necessary. 
As an attempt to improve on RPA, the Higher-order Random Phase Approximation (HRPA) was derived, which employs a wavefunction correct through second order in fluctuation potential.\cite{HRPA}
However, HRPA was shown to perform considerably worse than RPA. \cite{Haase,8RPAHRPASOPPA,11RPAHRPASOPPA} 
When also including double excitation- and de-excitation operators, one obtains the Second-Order Polarization Propagator (SOPPA), which is correct through second order.\cite{Oddershede, spas188} 
To improve the performance of RPA and HRPA while maintaining a computational cost lower than that of SOPPA, two new methods were developed: RPA(D) and HRPA(D).\cite{spas025, spas177, Haase} In SOPPA, double excitation- and de-excitation operators are added in the beginning, so they are also included in the iterative process. In RPA(D) and HRPA(D), however, the polarizabilities are first calculated using only RPA and HRPA, respectively, with only single excitation- and de-excitation operators. Then, using pseudo-perturbation theory,\cite{Book} a correction with double excitation- and de-excitation operators is added. This is called a doubles correction. Because the doubles correction is non-iterative, RPA(D) and HRPA(D), while attempting to approximate SOPPA results, are much less computationally demanding than SOPPA.

SOPPA-based methods have already been benchmarked for polarizabilities with various types of molecules.\cite{1SOPPA, 3Correlatedpropagator, 2Dipole, 4SOPPACCSD, 5Staticdynamic, 6Linearresponse, 8RPAHRPASOPPA, 9SOPPA, 10RPAHRPASOPPA, 11RPAHRPASOPPA}  
However, in the present paper, a new set of 43 molecules, presented by Hickey and Rowley,\cite{Hickey} including both non-aromatic- and aromatic molecules, will be studied. Previous benchmark studies have studied non-aromatic molecules \cite{1SOPPA, 3Correlatedpropagator, 2Dipole, 4SOPPACCSD, 5Staticdynamic, 6Linearresponse} and aromatic molecules \cite{8RPAHRPASOPPA,9SOPPA,10RPAHRPASOPPA} separately. Jørgensen \cite{8RPAHRPASOPPA} found that, opposed to previous calculations for small, non-aromatic molecules,\cite{1SOPPA, 3Correlatedpropagator, 2Dipole, 4SOPPACCSD, 5Staticdynamic, 6Linearresponse} RPA had an excellent performance for aromatic molecules. One study included both non-aromatic and aromatic molecules,\cite{11RPAHRPASOPPA} but without focusing on aromaticity. Therefore, this study will compare the performances of the aforementioned methods on non-aromatic- and aromatic molecules with CCSD results as reference data, which are both well-known methods. Furthermore, results from all computational methods will be compared to experimental data.

The structure of this work is as follows. Section 2 offers a brief overview of the theoretical foundations of the RPA(D) and HRPA(D) approximations. Section 3 outlines the computational methods employed in our study. In Section 4, we present our findings, distinguishing between static and frequency‑dependent polarizabilities and between aromatic and non‑aromatic molecules. We also compare our results with earlier theoretical and experimental data. Concluding remarks are provided in the final section.

\section{Theory}

The polarizability $\boldsymbol{\alpha}$ is a symmetric $3 \times 3$ tensor with six independent elements \cite{Book}:

\begin{equation}
    \boldsymbol{\alpha} = \begin{pmatrix}
        \alpha_{xx} & \alpha_{xy} & \alpha_{xz} \\
        \alpha_{yx} & \alpha_{yy} & \alpha_{yz} \\
        \alpha_{zx} & \alpha_{zy} & \alpha_{zz}
    \end{pmatrix}
\end{equation}

In this paper, however, only the isotropic polarizabilities $\alpha$ will be reported.

\begin{equation}
    \alpha = \frac{1}{3} \left(\alpha_{xx} + \alpha_{yy} + \alpha_{zz}\right)
\end{equation}

Traditionally, using exact state perturbation theory, the frequency dependent polarizability is expressed as a sum over the exact eigenstates of the molecule. However, since this requires knowledge of all excited states, it is usually calculated as a polarization propagator/linear response function \cite{Polprop} $\left<\left<\hat{\mu}_\alpha ; \, \hat{\mu}_\beta \right>\right>_\omega$:

\begin{equation}
    \alpha_{\alpha \beta}(\omega) = - \left<\left<\hat{\mu}_\alpha ; \, \hat{\mu}_\beta \right>\right>_\omega
\end{equation}
where the dipole moment operator in direction $\alpha$, $\hat{\mu}_\alpha$, is defined like as \cite{Book}:

\begin{equation}
    \hat{\mu}_\alpha = \sum_i q_i R_{i,\alpha}
\end{equation}
where $q_i$ and $R_{i,\alpha}$ are the charge and the $\alpha$ component of the position vector of particle $i$, respectively.

When approximating the polarization propagator at the TD-HF/RPA level, one gets the following \cite{Book}:
\begin{equation}
    \left<\left<\hat{\mu}_\alpha; \, \hat{\mu}_{\beta}\right>\right>_\omega^{\text{RPA}} = \begin{pmatrix}
        ^e\widetilde{\boldsymbol{\mu}}_\alpha^{(0)} & ^d\widetilde{\boldsymbol{\mu}}_\alpha^{(0)}
    \end{pmatrix} \begin{pmatrix}
        ^e \mathbf{X}_\beta^{\text{RPA}} \\
        ^d \mathbf{X}_\beta^{\text{RPA}}
    \end{pmatrix}
\end{equation}
with the RPA solution vector defined as
\begin{equation}
    \begin{pmatrix}
        ^e \mathbf{X}_\beta^{\text{RPA}} \\
        ^d \mathbf{X}_\beta^{\text{RPA}}
    \end{pmatrix} = \begin{pmatrix}
        \omega \mathbf{1} - \mathbf{A}^{(0, 1)} & -\mathbf{B}^{(1)} \\
        -\mathbf{B}^{(1)} & -\omega \mathbf{1} - \mathbf{A}^{(0, 1)}
    \end{pmatrix}^{-1} \begin{pmatrix}
        ^e \boldsymbol{\mu}_\beta^{(0)} \\
        ^d \boldsymbol{\mu}_\beta^{(0)}
    \end{pmatrix}
\end{equation}

The solution vector is thus obtained iteratively as the solution of the inhomogeneous set of linear response equations:
\begin{equation}
    \begin{pmatrix}
        \omega \mathbf{1} - \mathbf{A}^{(0, 1)} & -\mathbf{B}^{(1)} \\
        -\mathbf{B}^{(1)} & -\omega \mathbf{1} - \mathbf{A}^{(0, 1)}
    \end{pmatrix} \begin{pmatrix}
        ^e \mathbf{X}_\beta^{\text{RPA}} \\
        ^d \mathbf{X}_\beta^{\text{RPA}}
    \end{pmatrix} = \begin{pmatrix}
        ^e \boldsymbol{\mu}_\beta^{(0)} \\
        ^d \boldsymbol{\mu}_\beta^{(0)}
    \end{pmatrix}
    \label{eq.leqRPA}
\end{equation}

When the doubles correction is added to yield RPA(D), the polarization propagator becomes the sum of the RPA polarization propagator and two corrections:\cite{spas177}
\begin{equation}
    \left<\left<\hat{\mu}_\alpha; \, \hat{\mu}_{\beta}\right>\right>_\omega^{\text{RPA(D)}} = \left<\left<\hat{\mu}_\alpha; \, \hat{\mu}_{\beta}\right>\right>_\omega^{\text{RPA}} + \left<\left<\hat{\mu}_\alpha; \, \hat{\mu}_{\beta}\right>\right>_\omega^{\text{corr,RPA}}
\end{equation}
one with the second-order Møller-Plesset perturbation theory\cite{MP, HRPA} corrections to the single excitations and de-excitations, $\mathbf{A}^{(2)}$, $\mathbf{B}^{(2)}$, $\mathbf{\Sigma}^{(2)}$ and $\boldsymbol{\mu}_\beta^{(2)}$, and one with double excitation- and de-excitation operators:
\begin{equation}
    \left<\left<\hat{\mu}_\alpha; \, \hat{\mu}_{\beta}\right>\right>_\omega^{\text{corr,RPA}} = \left<\left<\hat{\mu}_\alpha; \, \hat{\mu}_{\beta}\right>\right>_\omega^{\text{corr,S}} + \left<\left<\hat{\mu}_\alpha; \, \hat{\mu}_{\beta}\right>\right>_\omega^{\text{corr,D}}
\end{equation}
defined as
\begin{eqnarray}
    \left<\left<\hat{\mu}_\alpha; \, \hat{\mu}_{\beta}\right>\right>_\omega^{\text{corr,S}} 
    &=& \begin{pmatrix}
        ^e \widetilde{\mathbf{X}}^{\text{RPA}}_\alpha &  ^d \widetilde{\mathbf{X}}^{\text{RPA}}_\alpha
    \end{pmatrix} 
    \begin{pmatrix}
        ^e \boldsymbol{\mu}_\beta^{(2)} \\
        ^d \boldsymbol{\mu}_\beta^{(2)}
    \end{pmatrix} \nonumber \\
    &&+ \begin{pmatrix}
        ^e \widetilde{\mathbf{X}}^{\text{RPA}}_\alpha &  ^d \widetilde{\mathbf{X}}^{\text{RPA}}_\alpha
    \end{pmatrix} 
    \begin{pmatrix}
        \mathbf{A}^{(2)} - \omega \mathbf{\Sigma}^{(2)} & \mathbf{B}^{(2)} \\
        \mathbf{B}^{(2)} & \mathbf{A}^{(2)} + \omega \boldsymbol{\Sigma}^{(2)}
    \end{pmatrix} 
    \begin{pmatrix}
        ^e \mathbf{X}_\beta^{\text{RPA}} \\
        ^d \mathbf{X}_\beta^{\text{RPA}}
    \end{pmatrix} \nonumber \\
    &&+ \begin{pmatrix}
        ^e\widetilde{\boldsymbol{\mu}}^{(2)}_\alpha & ^d\widetilde{\boldsymbol{\mu}}^{(2)}_\alpha
    \end{pmatrix} 
    \begin{pmatrix}
        ^e \mathbf{X}_\beta^{\text{RPA}} \\
        ^d \mathbf{X}_\beta^{\text{RPA}}
    \end{pmatrix}
\end{eqnarray}    
and
\begin{equation}
\begin{split}
    \left<\left<\hat{\mu}_\alpha; \, \hat{\mu}_{\beta}\right>\right>_\omega^{\text{corr,D}} = -\left[\begin{pmatrix}
        ^e\widetilde{\boldsymbol{\Pi}}^{(1)}_\alpha & ^d\widetilde{\boldsymbol{\Pi}}^{(1)}_\alpha
    \end{pmatrix} 
    + \begin{pmatrix}
        ^e \widetilde{\mathbf{X}}^{\text{RPA}}_\alpha &  ^d \widetilde{\mathbf{X}}^{\text{RPA}}_\alpha
    \end{pmatrix} 
    \begin{pmatrix}
        \widetilde{\mathbf{C}}^{(1)} & \mathbf{0} \\
        \mathbf{0} & \widetilde{\mathbf{C}}^{(1)}
    \end{pmatrix}\right] \times
    \\
    \begin{pmatrix}
        \mathbf{D}^{(0)} - \omega \mathbf{1} & \mathbf{0} \\
        \mathbf{0} & \mathbf{D}^{(0)} + \omega \mathbf{1}
    \end{pmatrix}^{-1} \left[\begin{pmatrix}
        ^e\mathbf{\Pi}_\beta^{(1)} \\
        ^d\mathbf{\Pi}_\beta^{(1)}
    \end{pmatrix} + \begin{pmatrix}
        \mathbf{C}^{(1)} & \mathbf{0} \\
        \mathbf{0} & \mathbf{C}^{(1)}
    \end{pmatrix} \begin{pmatrix}
        ^e \mathbf{X}_\beta^{\text{RPA}} \\
        ^d \mathbf{X}_\beta^{\text{RPA}}
    \end{pmatrix} \right]
\end{split}
\end{equation}

Since the solution vector from just RPA is used, the correction $\left<\left<\hat{\mu}_\alpha; \, \hat{\mu}_{\beta}\right>\right>_\omega^{\text{corr}}$ is obtained non-iteratively and thus at low additional computational cost.

In HRPA,\cite{HRPA} the polarization propagator is obtained like as:
\begin{equation}
    \left<\left<\hat{\mu}_\alpha; \, \hat{\mu}_{\beta}\right>\right>_\omega^{\text{HRPA}} = \begin{pmatrix}
        ^e\widetilde{\boldsymbol{\mu}}_\alpha^{(0, 2)} & 
        ^d \widetilde{\boldsymbol{\mu}}_\alpha^{(0, 2)}
    \end{pmatrix} \begin{pmatrix}
        ^e \mathbf{X}^{\text{HRPA}}_\beta \\
        ^d \mathbf{X}^{\text{HRPA}}_\beta
    \end{pmatrix}
\end{equation}
where the HRPA solution vector is defined as
\begin{equation}
    \begin{pmatrix}
        ^e \mathbf{X}^{\text{HRPA}}_\beta \\
        ^d \mathbf{X}^{\text{HRPA}}_\beta
    \end{pmatrix} = \begin{pmatrix}
        \omega \left(\mathbf{1} + \boldsymbol{\Sigma}^{(2)}\right) - \mathbf{A}^{(0, 1, 2)} & -\mathbf{B}^{(1,2)} \\
        -\mathbf{B}^{(1,2)} & -\omega \left(\mathbf{1} + \boldsymbol{\Sigma}^{(2)}\right) - \mathbf{A}^{(0, 1, 2)}
    \end{pmatrix}^{-1} \begin{pmatrix}
        ^e \boldsymbol{\mu}_\beta^{(0, 2)} \\
        ^d \boldsymbol{\mu}_\beta^{(0, 2)}
    \end{pmatrix}
        \label{eq.leqHRPA}
\end{equation}

When adding the doubles correction, the polarization propagator $\left<\left<\hat{\mu}_\alpha; \, \hat{\mu}_{\beta}\right>\right>_\omega^{\text{HRPA(D)}}$ becomes \cite{Haase, 8RPAHRPASOPPA}
with
\begin{equation}
    \left<\left<\hat{\mu}_\alpha; \, \hat{\mu}_{\beta}\right>\right>_\omega^{\text{HRPA(D)}} = \left<\left<\hat{\mu}_\alpha; \, \hat{\mu}_{\beta}\right>\right>_\omega^{\text{HRPA}} + \left<\left<\hat{\mu}_\alpha; \, \hat{\mu}_{\beta}\right>\right>_\omega^{\text{corr,HRPA}}
\end{equation}
with the doubles correction defined similiarly to RPA(D) but here with the HRPA solution vector
\begin{equation}
\begin{split}
    \left<\left<\hat{\mu}_\alpha; \, \hat{\mu}_{\beta}\right>\right>_\omega^{\text{corr,HRPA}} = - \left[\begin{pmatrix}
        ^e\widetilde{\boldsymbol{\Pi}}^{(1)}_\alpha & ^d\widetilde{\boldsymbol{\Pi}}^{(1)}_\alpha
    \end{pmatrix} 
    + \begin{pmatrix}
        ^e\widetilde{\mathbf{X}}_\alpha^{\text{HRPA}} & ^d\widetilde{\mathbf{X}}_\alpha^{\text{HRPA}}
    \end{pmatrix} 
    \begin{pmatrix}
        \widetilde{\mathbf{C}}^{(1)} & \mathbf{0} \\
        \mathbf{0} & \widetilde{\mathbf{C}}^{(1)}
    \end{pmatrix}\right] \times
    \\
    \begin{pmatrix}
        \mathbf{D}^{(0)} - \omega \mathbf{1} & \mathbf{0} \\
        \mathbf{0} & \mathbf{D}^{(0)} + \omega \mathbf{1}
    \end{pmatrix}^{-1} \left[\begin{pmatrix}
        ^e \boldsymbol{\Pi}^{(1)}_\beta \\
        ^d \boldsymbol{\Pi}^{(1)}_\beta
    \end{pmatrix} + \begin{pmatrix}
        \mathbf{C}^{(1)} & \mathbf{0} \\
        \mathbf{0} & \mathbf{C}^{(1)}
    \end{pmatrix} \begin{pmatrix}
        ^e \mathbf{X}^{\text{HRPA}}_\beta \\
        ^d \mathbf{X}^{\text{HRPA}}_\beta
    \end{pmatrix}\right]
\end{split}
\end{equation}

Finally, in the original second order polarization propagator approximation (SOPPA)\cite{Oddershede, spas188} the polarization propagator is calculated as:
\begin{equation}
    \left<\left<\hat{\mu}_\alpha; \, \hat{\mu}_{\beta}\right>\right>_\omega^{\text{SOPPA}} = \begin{pmatrix}
        ^e\widetilde{\boldsymbol{\mu}}_\alpha^{(0,2)} & ^d\widetilde{\boldsymbol{\mu}}_\alpha^{(0,2)} & ^e\widetilde{\boldsymbol{\Pi}}_\alpha^{(1)} & ^d\widetilde{\boldsymbol{\Pi}}_\alpha^{(1)}
    \end{pmatrix} \begin{pmatrix}
        ^e\mathbf{X}^{\text{SOPPA}}_\beta \\
        ^d\mathbf{X}^{\text{SOPPA}}_\beta \\
        ^e\boldsymbol{\Xi}^{\text{SOPPA}}_\beta \\
        ^d\boldsymbol{\Xi}^{\text{SOPPA}}_\beta
    \end{pmatrix}
\end{equation}
with the SOPPA solution vector defined as
\begin{equation}
\begin{split}
    \begin{pmatrix}
        ^e\mathbf{X}^{\text{SOPPA}}_\beta \\
        ^d\mathbf{X}^{\text{SOPPA}}_\beta \\
        ^e\boldsymbol{\Xi}^{\text{SOPPA}}_\beta \\
        ^d\boldsymbol{\Xi}^{\text{SOPPA}}_\beta
    \end{pmatrix} 
    = \begin{pmatrix}
        \omega \boldsymbol{\Sigma}^{(0,2)} - \mathbf{A}^{(0, 1, 2)} & -\mathbf{B}^{(1,2)} & -\mathbf{\widetilde{C}}^{(1)} & \mathbf{0} \\
        -\mathbf{B}^{(1,2)} & -\omega \boldsymbol{\Sigma}^{(0,2)} -\mathbf{A}^{(0, 1, 2)} & \mathbf{0} & -\mathbf{\widetilde{C}}^{(1)} \\
        -\mathbf{C}^{(1)} & \mathbf{0} & \omega - \mathbf{D}^{(0)} & \mathbf{0} \\
        \mathbf{0} & -\mathbf{C}^{(1)} & \mathbf{0} & -\omega - \mathbf{D}^{(0)}
    \end{pmatrix}^{-1} 
    \begin{pmatrix}
        ^e \boldsymbol{\mu}_\beta^{(0,2)} \\
        ^d \boldsymbol{\mu}_\beta^{(0,2)} \\
        ^e \boldsymbol{\Pi}_\beta^{(1)} \\
        ^d \boldsymbol{\Pi}_\beta^{(1)}
    \end{pmatrix}
\end{split}
    \label{eq.leqSOPPA}
\end{equation}

The different submatrices and vectors are defined as\cite{Book}
\begin{equation}
    \mathbf{A}^{(0,1,2)}_{ai,bj} = \left<\Phi_0^{\text{MP}} \left|\left[q_{ai}, \left[\hat{F} + \hat{V}, q_{bj}^\dagger\right]\right]\right|\Phi_0^{\text{MP}}\right>^{(0,1,2)}
\end{equation}
\begin{equation}
    \mathbf{B}^{(1,2)}_{ai,bj} = \left<\Phi_0^{\text{MP}}\left|\left[q_{ai}, \left[\hat{F} + \hat{V}, q_{bj}\right]\right]\right|\Phi_0^{\text{MP}}\right>^{(1,2)}
\end{equation}
\begin{equation}
    \mathbf{C}^{(1)}_{aibj,ck} = \left<\Phi_0^{\text{MP}}\left|\left[q_{ai} q_{bj}, \left[\hat{F} + \hat{V}, q_{ck}^\dagger\right]\right]\right|\Phi_0^{\text{MP}}\right>^{(1,2)}
\end{equation}
\begin{equation}
    \boldsymbol{\Sigma}^{(0,2)}_{ai,bj} = \left<\Phi_0^{\text{MP}}\left|\left[q_{ai}, q_{bj}^\dagger\right]\right|\Phi_0^{\text{MP}}\right>^{(0,2)}
\end{equation}
\begin{equation}
    ^e \boldsymbol{\mu}^{(0,2)}_{\alpha,ai} = \left<\Phi_0^{\text{MP}}\left|\left[q_{ai}, \hat{\mu}_\alpha\right]\right|\Phi_0^{\text{MP}}\right>^{(0,2)}
\end{equation}
\begin{equation}
    ^d \boldsymbol{\mu}^{(0,2)}_{\alpha,ai} = \left<\Phi_0^{\text{MP}}\left|\left[q_{ai}^\dagger, \hat{\mu}_\alpha\right]\right|\Phi_0^{\text{MP}}\right>^{(0,2)}
\end{equation}
\begin{equation}
    ^e \boldsymbol{\Pi}^{(1)}_{\alpha,aibj} = \left<\Phi_0^{\text{MP}}\left|\left[q_{ai}q_{bj}, \hat{\mu}_\alpha\right]\right|\Phi_0^{\text{MP}}\right>^{(1)}
\end{equation}
\begin{equation}
    ^d \boldsymbol{\Pi}^{(1)}_{\alpha,aibj} = \left<\Phi_0^{\text{MP}}\left|\left[q_{ai}^\dagger q_{bj}^\dagger, \hat{\mu}_\alpha\right]\right|\Phi_0^{\text{MP}}\right>^{(1)}
\end{equation}
where $\left|\Phi_0^{\text{MP}}\right>$ is the Møller-Plesset perturbation theory wavefunction\cite{MP} and $q_{ai}^\dagger$ and $q_{ai}^\dagger q_{bj}^\dagger$ are spin-adapted single and double excitation operators.\cite{ci777pac, mbpt89pp, jod130}

Using CC2 or CCSD singles and doubles amplitudes instead of the Møller-Plesset correlation coefficients in the SOPPA equations, one obtains two approaches, which are called SOPPA(CC2)\cite{spas095} and SOPPA(CCSD), \cite{4SOPPACCSD} respectively.

To calculate excitation energies, $\omega_n$, with all previously mentioned methods, the RPA, HRPA or SOPPA generalized eigenvalue equations\cite{Excen,Book} have to be solved instead of the linear response equations in eqs. \ref{eq.leqRPA}, \ref{eq.leqHRPA} and \ref{eq.leqSOPPA}.
For RPA(D), the excitation energy is then given as \cite{spas025,Haase}
\begin{equation}
    \omega^{\text{RPA(D)}}_n = \omega^{\text{RPA}}_n + \omega^{\text{corr,RPA}}_n
\end{equation}
where $\omega^{\text{RPA}}_n$ is the excitation energy at the RPA level and the corrections are obtained as
\begin{equation}
    \omega^{\text{corr,RPA}}_n = \omega^{\text{corr,S}}_n + \omega^{\text{corr,D}}_n
\end{equation}
with
\begin{equation}
    \begin{split}
        \omega^{\text{corr,S}}_n = \begin{pmatrix}
            ^e\widetilde{\mathbf{X}}_n^{\text{RPA}} & ^d\widetilde{\mathbf{X}}_n^{\text{RPA}}
        \end{pmatrix} 
        \begin{pmatrix}
            \mathbf{A}^{(2)} - \omega_n^{\text{RPA}} \boldsymbol{\Sigma}^{(2)} & \mathbf{B}^{(2)} \\
            \mathbf{B}^{(2)} & \mathbf{A}^{(2)} + \omega_n^{\text{RPA}} \boldsymbol{\Sigma}^{(2)}
        \end{pmatrix} 
        \begin{pmatrix}
            ^e\mathbf{X}_n^{\text{RPA}} \\
            ^d\mathbf{X}_n^{\text{RPA}}
        \end{pmatrix}
    \end{split}
\end{equation}
and
\begin{equation}
    \begin{split}
        \omega^{\text{corr,D}}_n = -\begin{pmatrix}
            ^e\widetilde{\mathbf{X}}_n^{\text{RPA}} & ^d\widetilde{\mathbf{X}}_n^{\text{RPA}}
        \end{pmatrix} 
        \begin{pmatrix}
            \widetilde{\mathbf{C}}^{(1)} & 0 \\
            0 & \widetilde{\mathbf{C}}^{(1)}
        \end{pmatrix} 
        \begin{pmatrix}
            \mathbf{D}^{(0)} - \omega_n^{\text{RPA}} & 0 \\
            0 & \mathbf{D}^{(0)} + \omega_n^{\text{RPA}}
        \end{pmatrix}^{-1} \times
        \\
        \begin{pmatrix}
            \mathbf{C}^{(1)} & 0 \\
            0 & \mathbf{C}^{(1)}
        \end{pmatrix} \begin{pmatrix}
            ^e\mathbf{X}_n^{\text{RPA}} \\
            ^d\mathbf{X}_n^{\text{RPA}}
        \end{pmatrix}
    \end{split}
\end{equation}
Since only the RPA eigenvectors, $\begin{pmatrix}             ^e\widetilde{\mathbf{X}}_n^{\text{RPA}} & ^d\widetilde{\mathbf{X}}_n^{\text{RPA}} \end{pmatrix}$, are necessary, as with the calculation of the polarizability, the correction to the RPA excitation energy is non-iterative. 

The HRPA(D) excitation energy is calculated like as \cite{Haase}:
\begin{equation}
    \omega_n^{\text{HRPA(D)}} = \omega_n^{\text{HRPA}} + \omega_n^{\text{corr, HRPA}}
\end{equation}
with the doubles correction defined as
\begin{equation}
    \begin{split}
        \omega_n^{\text{corr, HRPA}} = \begin{pmatrix}
            ^e\widetilde{\mathbf{X}}_n^{\text{HRPA}} & ^d\widetilde{\mathbf{X}}_n^{\text{HRPA}}
        \end{pmatrix} \begin{pmatrix}
            \mathbf{A}^{(2)} - \omega_n^{\text{HRPA}} \boldsymbol{\Sigma}^{(2)} & \mathbf{B}^{(2)} \\
            \mathbf{B}^{(2)} & \mathbf{A}^{(2)} + \omega_n^{\text{HRPA}} \boldsymbol{\Sigma}^{(2)}
        \end{pmatrix} 
        \begin{pmatrix}
            ^e\mathbf{X}_n^{\text{HRPA}} \\
            ^d\mathbf{X}_n^{\text{HRPA}}
        \end{pmatrix}
    \end{split}
\end{equation}

\section{\label{Computational}Computational details}
The optimized geometries of the molecules were obtained using the MP2 method with the aug-cc-pVTZ basis set\cite{bs890115d, bs930115wd, bs940215wd} using the Gaussian program.\cite{g16} 
The benchmarking is done for static polarizabilities and dynamic polarizabilities at the wavelengths $355.0 \, \mathrm{nm}$ and $589.3 \, \mathrm{nm}$ with the RPA \cite{RPA}, RPA(D) \cite{Haase}, HRPA \cite{HRPA}, HRPA(D) \cite{Haase}, and SOPPA \cite{1SOPPA} methods using the Dalton program.\cite{dalton} 
CCSD results for polarizabilities were obtained with the CFOUR program \cite{cfourarticle} and with the Dalton program \cite{dalton} (Table \ref{SI-tab:Program} in the SI). CCSD results at wavelengths other than $355.0 \, \mathrm{nm}$ and $589.3 \, \mathrm{nm}$ for comparison with experimental data were obtained with the Daltion program.\cite{dalton}
All excitation energies have been calculated using the Dalton program. 

All the aforementioned methods have been used to calculate the polarizabilities of the 45 molecules presented by Hickey and Rowley.\cite{Hickey} The set of molecules contains 11 aromatic molecules (imidazole, furan, thiophene, chlorobenzene, toluene, phenol, benzene, fluorobenzene, pyridine, pyrrole, and pyrazole), one non-aromatic molecule with conjugated double bonds (1,3-butadiene), 19 non-aromatic organic molecules (ethanol, acetonitrile, fluoromethane, dimethyl sulfide, dimethyl ether, dimethylamine, trimethyl amine, dimethyl sulfone, ethene, propane, isubutene, 1-pentene, acetone, acetaldehyde, acetic acid, methyl formate, methyl acetate, cytosine, and methyl acetamide), seven diatomic molecules (nitrogen monoxide, chlorine, bromine, sulfur monoxide, oxygen, carbon monoxide, and nitrogen), two inorganic carbonyl compounds (carbon dioxide and sulfur dioxide), and five small inorganic molecules (hydrogen sulfide, ammonia, phosphine, water, and silane). The molecules $\mathrm{O_2}$ and $\mathrm{SO}$ have been removed because they are triplets, $\mathrm{NO}$ has been removed because it is a radical, and 1-pentene has been removed because of computational difficulties with having the calculations converge.
Experimental data for static polarizabilities were taken from Hickey and Rowley. \cite{Hickey}

\section{\label{Results}Results and discussion}
To benchmark SOPPA-based methods, the performances of RPA(D) and HRPA(D) (Tables \ref{SI-tab:datastatRPA}-\ref{SI-tab:data355RPA}  in the SI) as well as SOPPA, SOPPA(CC2), and SOPPA(CCSD) (Tables \ref{SI-tab:datastaticSOPPA}-\ref{SI-tab:data355SOPPA} in the SI) are compared to the performances of RPA and HRPA (Tables \ref{SI-tab:datastatRPA}-\ref{SI-tab:data355RPA} in the SI). CCSD has been shown to have an excellent performance, \cite{CCpol, CCbench} and CCSD results will therefore be used as reference data (Tables \ref{SI-tab:ccsdstatic}-\ref{SI-tab:ccsd355} in the SI). In the following, results obtained from the methods RPA, RPA(D), HRPA, HRPA(D), SOPPA, SOPPA(CC2), and SOPPA(CCSD) will be referred to as the \textit{calculations}. 
The statistical analyses in terms of the mean deviation (MD), the mean absolute deviation (MAD) and the standard deviation from the mean devation (StdDev) are made based on deviations of the calculations from the reference values:
\begin{equation}
    \mathrm{Dev.} = \alpha(\mathrm{calc.}) - \alpha(\mathrm{ref.})
\end{equation}
Therefore, a positive deviation corresponds to an overestimation compared to the reference data, and a negative deviation corresponds to an underestimation. For the static polarizabilities, experimental data for all molecules were provided.\cite{Hickey} Dynamic polarizabilities for furan and thiophene,\cite{Furanexp2,Furanexp3,Exp589} for water and carbon monoxide,\cite{PolCOwater} and for nitrogen \cite{PolN2} have been found. When examining the graphical representations of the deviations of the calculations from the reference values (Figures \ref{SI-fig:deviationfullout}-\ref{SI-fig:deviation355out} in the SI), it is evident that HRPA is the worst method by a large amount and systematically underestimates the polarizabilities, with high standard deviations. Therefore, HRPA has been removed in the following graphical representations.

Additionally, the deviations at the highest frequency (Figure \ref{SI-fig:deviation355out} in the SI) reveal data points for cytosine, bromine, and chlorine, whose calculated polarizabilities deviate significantly more from the CCSD results using the various methods.
Cytosine is responsible for the data points lying considerably higher than the remaining data points using RPA, SOPPA, SOPPA(CC2), and SOPPA(CCSD), and significantly lower using HRPA(D). Bromine yields the data point with a substantially higher deviation when using RPA(D). Chlorine yields the lowest-lying data points with SOPPA and SOPPA(CC2). The lowest singlet excitation energies of cytosine and bromine, contributing to their polarizability, have been calculated (Table \ref{SI-excbrcyt} in the SI). For cytosine, all the calculated excitation energies are higher than the highest frequency used in this paper to calculate the polarizabilities ($0.128347 \, \mathrm{au}$). However, SOPPA, SOPPA(CC2), and SOPPA(CCSD) yield excitation energies that lie sufficiently close to the highest frequency to introduce an error of significance. SOPPA(CC2) yields the highest excitation energy of the three methods, and it is observed in Figure \ref{SI-fig:deviation355out} that the error for this method is smaller than for SOPPA and SOPPA(CCSD). For bromine, only RPA and HRPA yield excitation energies higher than the highest frequency, with the remaining methods yielding lower excitation energies. The excitation energy from RPA is close enough to this frequency to introduce a significant error in the calculations. For chlorine, HRPA(D), SOPPA, and SOPPA(CC2) yield excitation energies lower than the frequency, and RPA(D) yields one higher, but very close. SOPPA(CCSD) and CCSD yield higher excitation energies, but close enough to introduce significant errors. Cytosine, bromine, and chlorine, therefore, will be removed from the data set when performing statistical analyses, which include the calculations at $355.0 \, \mathrm{nm}$.

In the statistical analyses, all methods will be ordered in terms of increasing accuracy and consistency, respectively. High accuracy will be defined as a low mean absolute deviation, and high consistency will be defined as a low standard deviation. Unless stated otherwise, an assessment of the performance of a method is based on both accuracy and consistency.
\subsection{Overall performance}
\begin{table}[h!]
\caption{Deviations of the polarizabilities calculated with various methods from the CCSD results (in au) averaged over all frequencies and all 41 molecules, with cytosine, bromine, and chlorine removed at the highest frequency.}
\label{tab:statanfull}
\centering
\begin{tabular}{lrrr}
\hline
\textbf{Method} & \textbf{MD} & \textbf{MAD} & \textbf{StdDev} \\ \hline
RPA             & $-1.57$                     & $1.80$                            & $1.63$                      \\
RPA(D)          & $0.96$                      & $0.99$                           & $0.94$                      \\
HRPA            & $-8.75$                     & $8.75$                           & $5.53$                      \\
HRPA(D)         & $-0.78$                     & $0.85$                           & $1.08$                      \\
SOPPA           & $1.69$                      & $1.70$                            & $1.36$                      \\
SOPPA(CC2)      & $1.54$                      & $1.56$                           & $1.25$                      \\
SOPPA(CCSD)     & $0.75$                      & $0.85$                           & $1.13$                      \\ \hline
\end{tabular}
\end{table}
\begin{figure}[h!]
    \centering
    \includegraphics[width=0.9\linewidth]{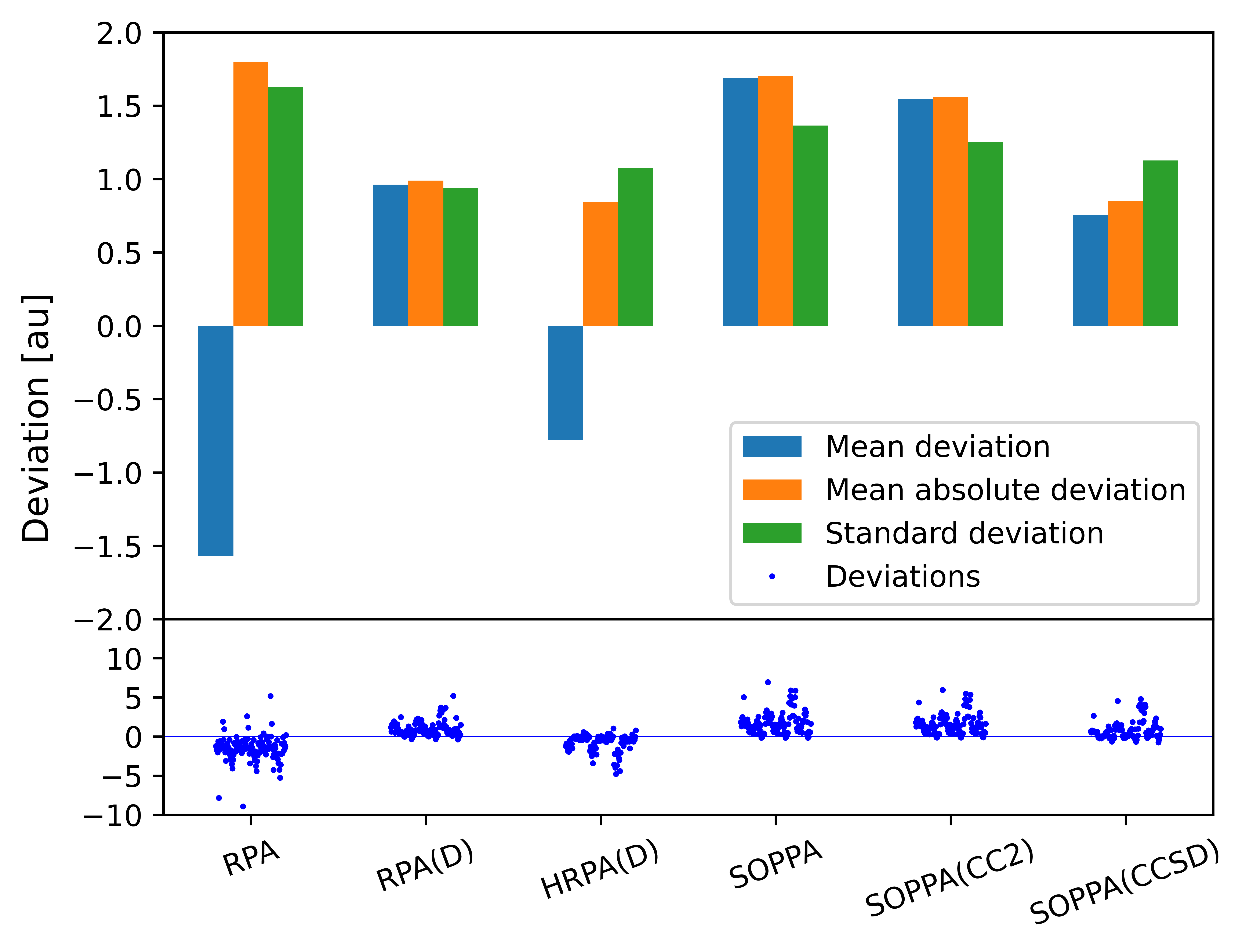}
    \caption{Deviations of the polarizabilities calculated with various methods from the CCSD results (in au) averaged over all frequencies and all 41 molecules, with cytosine, bromine, and chlorine removed at the highest frequency.}
    \label{fig:deviationfull}
\end{figure}
First, the results for all frequencies will be evaluated.
The statistical analysis of the full data set in Table \ref{tab:statanfull} and Figure \ref{fig:deviationfull} shows that, apart from HRPA, RPA has the worst performance, with the mean absolute- and standard deviations for SOPPA being 0.10 au lower and $0.27 \, \mathrm{au}$ lower, respectively, than for RPA. Using CC2 amplitudes in SOPPA(CC2) improves the performance compared to using MP2 correlation coefficients in SOPPA. The doubles correction in RPA(D) improves drastically on RPA, with a decrease in mean absolute deviation of $0.81 \, \mathrm{au}$ and a decrease in standard deviation of $0.69 \, \mathrm{au}$. RPA(D), therefore, also performs better than SOPPA(CC2). Likewise, the doubles correction in HRPA(D) improves drastically on HRPA. While HRPA(D) is more accurate than RPA(D), the standard deviation for HRPA(D) is $0.14 \, \mathrm{au}$ higher than that of RPA(D). The CCSD amplitudes in SOPPA(CCSD) improve on using CC2 amplitudes in SOPPA(CC2), and SOPPA(CCSD) is as accurate as HRPA(D), however, it is less consistent. The order of increasing accuracy is thus
\begin{equation*}
    \begin{split}
        \mathrm{HRPA < RPA < SOPPA < SOPPA(CC2) <
        RPA(D) < HRPA(D) /SOPPA(CCSD)}
    \end{split}
\end{equation*}
The order of increasing consistency is
\begin{equation*}
    \begin{split}
        \mathrm{HRPA < RPA < SOPPA < SOPPA(CC2) <
        SOPPA(CCSD) < HRPA(D) < RPA(D)}
    \end{split}
\end{equation*}

For a clearer picture of the performances of the methods, they will be assessed at the individual frequencies in the following.

\subsection{Static polarizabilities}
\begin{table}[h!]
\caption{Deviations of the static polarizabilities calculated with various methods from the CCSD results (in au) averaged over all 41 molecules.}
\label{tab:statanstatic}
\centering
\begin{tabular}{lrrr}
\hline
\textbf{Method} & \textbf{MD} & \textbf{MAD} & \textbf{StdDev} \\ \hline
RPA             & $-1.57$                     & $1.71$                           & $1.46$                      \\
RPA(D)          & $0.71$                      & $0.73$                           & $0.58$                      \\
HRPA            & $-7.87$                     & $7.87$                           & $4.70$                       \\
HRPA(D)         & $-0.42$                     & $0.48$                           & $0.57$                      \\
SOPPA           & $1.26$                      & $1.27$                           & $0.93$                      \\
SOPPA(CC2)      & $1.16$                      & $1.17$                           & $0.86$                      \\
SOPPA(CCSD)     & $0.30$                       & $0.43$                           & $0.55$                      \\ \hline
\end{tabular}
\end{table}

\begin{figure}[h!]
    \centering
    \includegraphics[width=0.9\linewidth]{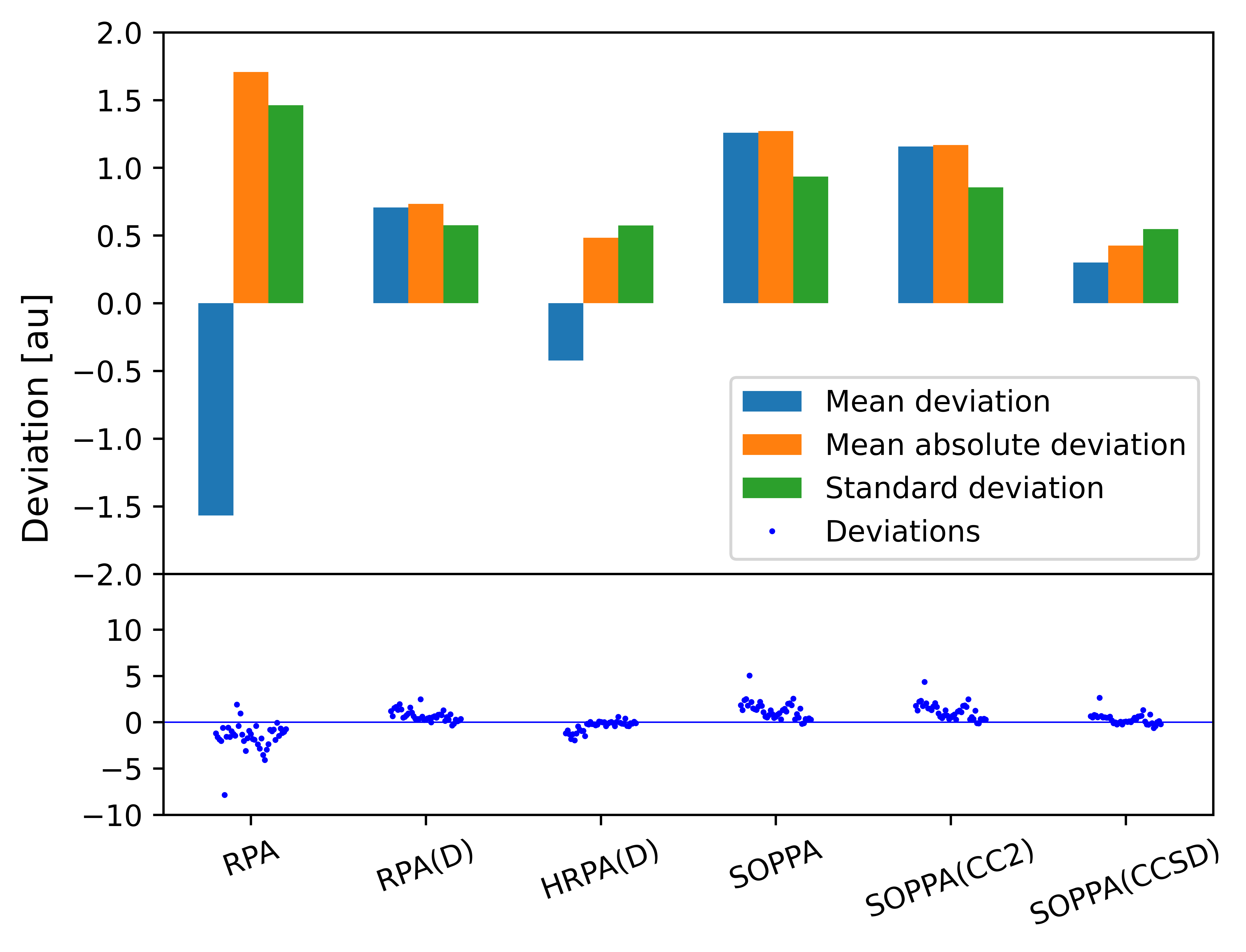}
    \caption{Deviations of the static polarizabilities calculated with various methods from the CCSD results (in au) averaged over all 41 molecules.}
    \label{fig:deviationstatic}
\end{figure}

The deviations of the calculated static polarizabilities from the CCSD results are analyzed in Table \ref{tab:statanstatic} and Figure \ref{fig:deviationstatic}.
The deviations show that RPA(D), SOPPA, SOPPA(CC2), and SOPPA(CCSD) all tend to overestimate the polarizabilities. Using CC2 amplitudes in SOPPA(CC2) is slightly better than using the MP2 correlation coefficients in SOPPA, but to such a small degree that it is doubtful that the increased computational cost is worth it. Both methods are better than RPA. When including the doubles correction in RPA(D), the performance drastically increases, and RPA(D) is better than both SOPPA and SOPPA(CC2). 
Although HRPA has by far the largest mean- and mean absolute deviation from the CCSD results, adding the doubles correction in HRPA(D) dramatically improves the results. 
HRPA(D) and SOPPA(CCSD) yield similar results, although HRPA(D) tends to underestimate the polarizabilities, whereas SOPPA(CCSD) tends to overestimate them. Both methods outperform the others, and SOPPA(CCSD) outperforms HRPA(D). The difference between SOPPA(CCSD) and HRPA(D) is sufficiently large for SOPPA(CCSD) to be used if high performance is important, but otherwise, HRPA(D) is likely the better choice.

\begin{figure}[h!]
    \centering
    \includegraphics[width=0.9\linewidth]{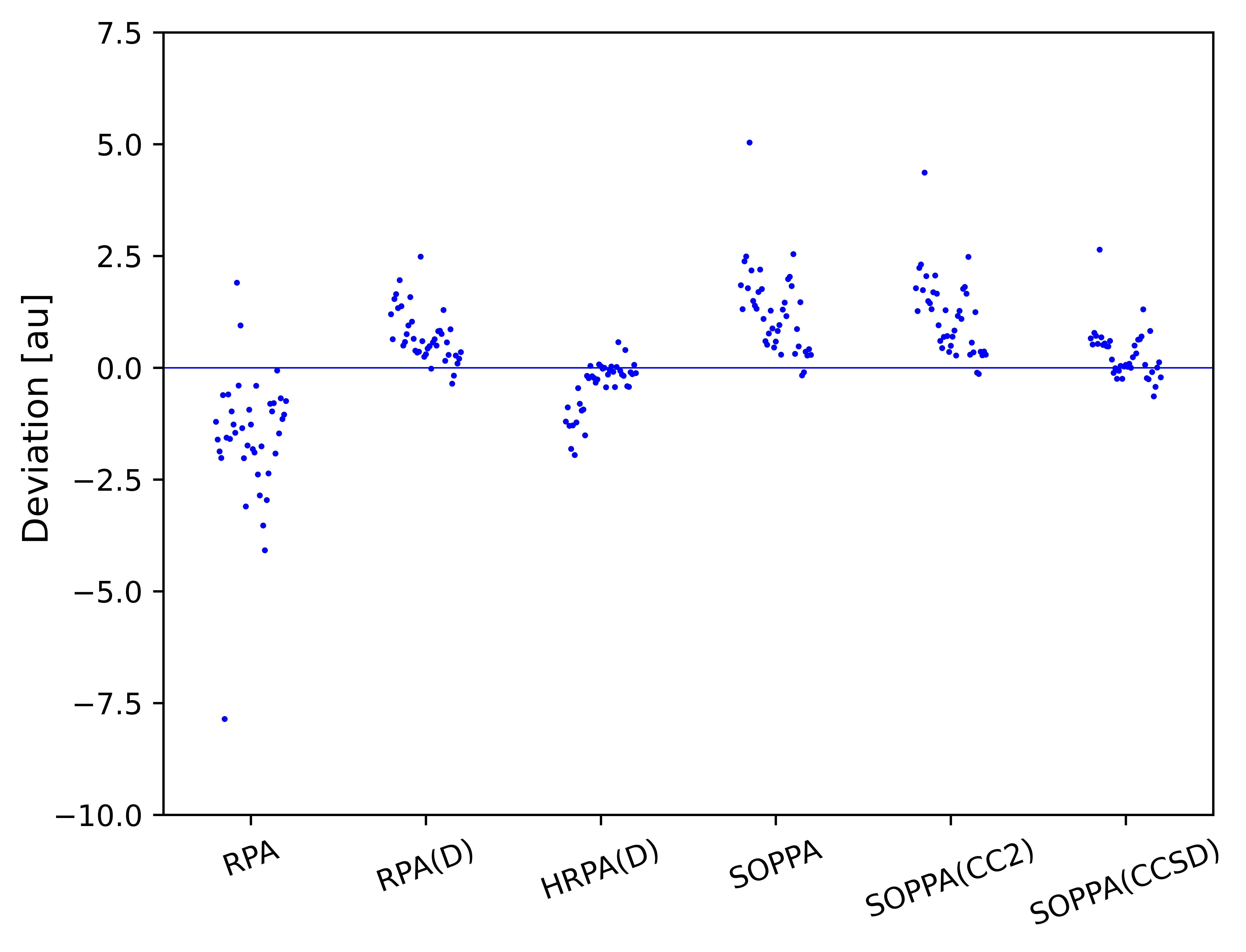}
    \caption{Deviations of the static polarizabilities of all 41 individual molecules calculated with various methods from the CCSD results (in au).}
    \label{fig:aromaticdev}
\end{figure}

The graphical representation of the deviations in Figure \ref{fig:deviationstatic} reveals 11 data points with significantly higher numerical deviations than the other molecules using HRPA(D) (Figure \ref{fig:aromaticdev}). 10 of the 11 data points stem from 10 of the 11 aromatic molecules (0.81-1.81 au), with furan having a slightly lower numerical deviation (0.46 au), though still higher than the remaining molecules. Cytosine has the highest numerical deviation of all molecules (1.95 au). One could argue that cytosine is aromatic because it has an aromatic resonance structure, however, it will be treated as a non-aromatic molecule in this paper. 
For RPA(D), the five aromatic molecules chlorobenzene, toluene, phenol, fluorobenzene, and benzene have the highest deviations, \textit{i.e.} $1.96 \, \mathrm{au}$, $1.65 \, \mathrm{au}$, $1.58 \, \mathrm{au}$, $1.54 \, \mathrm{au}$, $1.37 \, \mathrm{au}$, and $1.34 \, \mathrm{au}$, respectively, following trimethyl amine with $2.48 \, \mathrm{au}$ and cytosine with $1.96 \, \mathrm{au}$, and pyridine has a deviation of similar magnitude, \textit{i.e.} $1.20 \, \mathrm{au}$. 
The deviations of the remaining aromatic molecules, except furan, are also in the upper range, numerically (0.57-1.20 au). 
This tendency of the aromatic molecules to have deviations of a similar magnitude, but different from most non-aromatic molecules, cannot be seen with RPA, but is seen for SOPPA (1.31-2.49 au) and SOPPA(CC2) (1.27-2.31 au), and especially for SOPPA(CCSD) (0.48-0.78 au), their deviations are in the upper range.

\begin{table}[h!]
\caption{Deviations of the static polarizabilities calculated with various methods from the CCSD results (in au) averaged over the 30 non-aromatic molecules.}
\label{tab:statanstaticnonaromatic}
\centering
\begin{tabular}{lrrr}
\hline
\textbf{Method} & \textbf{MD} & \textbf{MAD} & \textbf{StdDev} \\ \hline
RPA         & $-1.65$            & $1.84$                  & $1.68$             \\
RPA(D)      & $0.56$             & $0.60$                   & $0.56$             \\
HRPA        & $-5.77$            & $5.77$                  & $3.32$             \\
HRPA(D)     & $-0.17$            & $0.25$                  & $0.40$              \\
SOPPA       & $1.05$             & $1.07$                  & $0.99$             \\
SOPPA(CC2)  & $0.93$             & $0.95$                  & $0.88$             \\
SOPPA(CCSD) & $0.19$             & $0.37$                  & $0.60$                       \\ \hline
\end{tabular}
\end{table}

\begin{figure}[h!]
    \centering
    \includegraphics[width=0.9\linewidth]{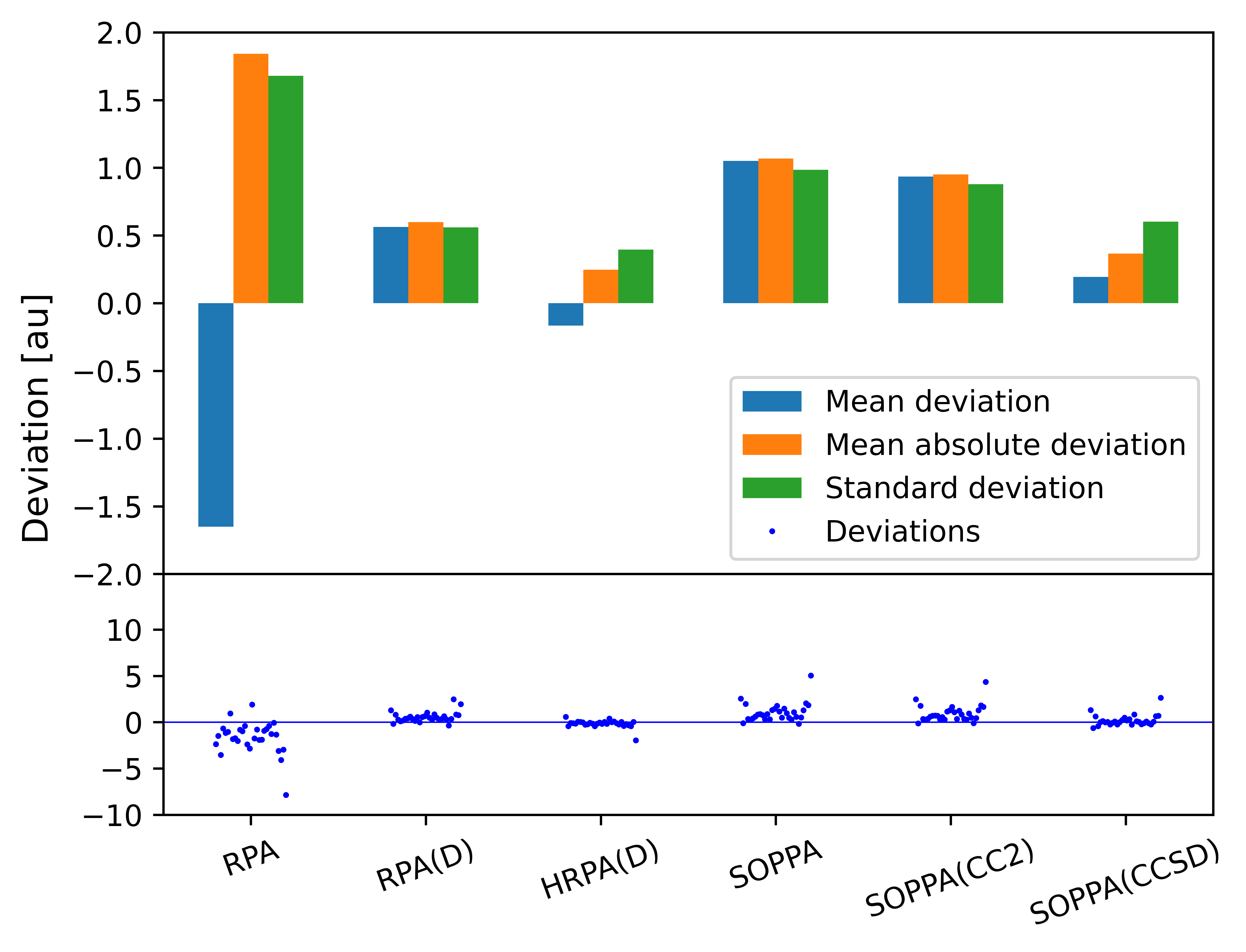}
    \caption{Deviations of the static polarizabilities calculated with various methods from the CCSD results (in au) averaged over the 30 non-aromatic molecules.}
    \label{fig:deviationstaticnonaromatic}
\end{figure}

To quantitatively investigate the effect of the aromatic molecules on the statistical data, they were removed from the data set, and a new statistical analysis was made for the non-aromatic molecules alone. 
Comparison of the new statistical data in Table \ref{tab:statanstaticnonaromatic} and Figure \ref{fig:deviationstaticnonaromatic} with that of the full data set in Table \ref{tab:statanstatic} and Figure \ref{fig:deviationstatic} shows that the inclusion of aromatic molecules in the data set is, as expected, not of much importance when using RPA compared to the other methods, with an increase in mean absolute deviation of $0.13 \, \mathrm{au}$ and an increase in standard deviation of $0.22 \, \mathrm{au}$. However, it makes a major difference for HRPA, with a decrease in mean absolute deviation of $2.10 \, \mathrm{au}$ and a decrease in standard deviation of $1.38 \, \mathrm{au}$. The difference is smaller, yet notable for the remaining methods. When the aromatic molecules are included, HRPA(D) has a lower numerical mean deviation and a lower mean absolute deviation than RPA(D), but a similar standard deviation. Removing the aromatic molecules (Table \ref{tab:statanstaticnonaromatic} and Figure \ref{fig:deviationstaticnonaromatic}) significantly increases the difference in performance between HRPA(D) and RPA(D) while simultaneously increasing the performance of both methods. Furthermore, HRPA(D) has the best performance of all methods. SOPPA(CCSD) exhibits a somewhat similar performance, with a numerical mean deviation almost identical to that of HRPA(D). 
However, in terms of the mean absolute deviation it is notably worse than HRPA(D) and has a standard deviation $1.5$ times as high as that of HRPA(D). SOPPA(CCSD) is more accurate than RPA(D), but it is slightly less consistent. SOPPA and SOPPA(CC2) perform worse than RPA(D), with SOPPA(CC2) being slightly better than SOPPA. All these methods yield better results than RPA. HRPA is, despite a much better performance than with the full data set, still the worst method by a large amount. When considering the non-aromatic molecules, the order of increasing accuracy is then
\begin{equation*}
    \begin{split}
        \mathrm{HRPA < RPA < SOPPA < SOPPA(CC2)
< RPA(D) < SOPPA(CCSD) < HRPA(D)}
    \end{split}
\end{equation*}
while the order of increasing consistency is
\begin{equation*}
    \begin{split}
        \mathrm{HRPA < RPA < SOPPA < SOPPA(CC2)
< SOPPA(CCSD) < RPA(D) < HRPA(D)}
    \end{split}
\end{equation*}

\begin{table}[h!]
\caption{Deviations of the static polarizabilities calculated with various methods from the CCSD results (in au) averaged over the 11 aromatic molecules.}
\label{tab:statanstaticaromatic}
\centering
\begin{tabular}{lrrr}
\hline
\textbf{Method} & \textbf{MD} & \textbf{MAD} & \textbf{StdDev} \\ \hline
RPA         & $-1.34$            & $1.34$                  & $0.44$             \\
RPA(D)      & $1.10$             & $1.10$                  & $0.41$             \\
HRPA        & $-13.58$           & $13.58$                 & $2.82$             \\
HRPA(D)     & $-1.13$            & $1.13$                  & $0.35$             \\
SOPPA       & $1.82$             & $1.82$                  & $0.41$             \\
SOPPA(CC2)  & $1.76$             & $1.76$                  & $0.35$             \\
SOPPA(CCSD) & $0.59$             & $0.59$                  & $0.10$                      \\ \hline
\end{tabular}
\end{table}

\begin{figure}[h!]
    \centering
    \includegraphics[width=0.9\linewidth]{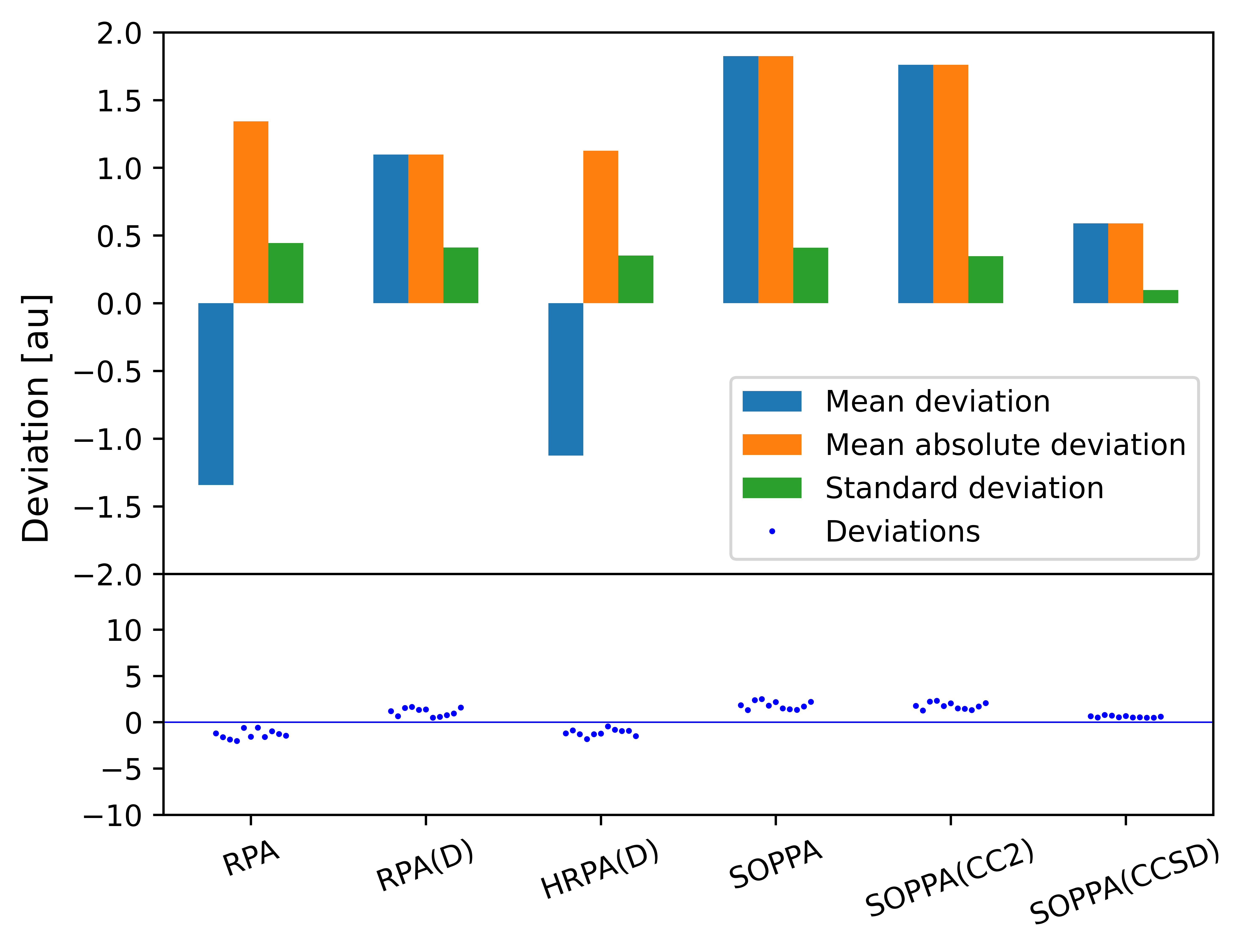}
    \caption{Deviations of the static polarizabilities calculated with various methods from the CCSD results (in au) averaged over the 11 aromatic molecules.}
    \label{fig:deviationstaticaromatic}
\end{figure}

Now, the aromatic molecules are considered, and a statistical analysis is performed on their deviations in Table \ref{tab:statanstaticaromatic} and Figure \ref{fig:deviationstaticaromatic}.
It is immediately observed that all methods are significantly worse at predicting the polarizabilities of aromatic molecules than of non-aromatic molecules. 
Interestingly, for all methods, the numerical mean deviation and the mean absolute deviation are the same when considering only the aromatic molecules, 
whereas they for all methods, except HRPA, are different for the non-aromatic molecules (Table \ref{tab:statanstaticnonaromatic} and Figure \ref{fig:deviationstaticnonaromatic}) and for the full data set (Table \ref{tab:statanstatic} and Figure \ref{fig:deviationstatic}).
This implies of course that the methods consistently over- or underestimate the CCSD results,
With all data sets (Tables \ref{tab:statanstatic}-\ref{tab:statanstaticaromatic} and Figures \ref{fig:deviationstatic}-\ref{fig:deviationstaticaromatic}), the mean deviation is negative for RPA, HRPA, and HRPA(D) and positive for the remaining methods. This indicates that the tendency of a method to over- or underestimate the polarizability of a molecule is almost certainly followed when the molecule is aromatic, but not necessarily when it is non-aromatic. This is confirmed by the graphical representation of the deviations of non-aromatic molecules (Figure \ref{fig:deviationstaticnonaromatic}). 

Using CCSD amplitudes in SOPPA(CCSD) yields the best results. RPA(D) has a lower mean absolute deviation than HRPA(D) by $0.03 \, \mathrm{au}$, but the standard deviation of HRPA(D) is $0.06 \, \mathrm{au}$ lower than that of RPA(D). They are therefore very close in performance, with RPA(D) overestimating the polarizabilities and HRPA(D) underestimating them. RPA(D) has a marginally better performance than HRPA(D) in terms of consistency, while HRPA(D) is marginally better in terms of accuracy. 
The doubles corrections in RPA(D) and HRPA(D) improve thus the results from their RPA and HRPA counterparts. RPA has a lower mean absolute deviation than SOPPA and SOPPA(CC2), but SOPPA and SOPPA(CC2) have lower standard deviations than RPA. SOPPA and SOPPA(CC2) are similar in performance, but using the CC2 amplitudes is slightly better than using the MP2 correlation coefficients. Consequently, the standard deviation for SOPPA is identical to that of RPA(D), whereas the standard deviation for SOPPA(CC2) is $0.06 \, \mathrm{au}$ lower than that of RPA(D). However, for SOPPA(CC2), the standard deviation is identical to that of HRPA(D). HRPA is, again, the worst method and drastically underestimates the polarizabilities. The order of increasing accuracy is thus
\begin{equation*}
    \begin{split}
        \mathrm{HRPA < SOPPA < SOPPA(CC2) < RPA < 
        HRPA(D) < RPA(D) < SOPPA(CCSD)}
    \end{split}
\end{equation*}
while the order of increasing consistency is
\begin{equation*}
    \begin{split}
        \mathrm{HRPA < RPA < RPA(D) / SOPPA < 
        HRPA(D) / SOPPA(CC2) < SOPPA(CCSD)}
    \end{split}
\end{equation*}

\subsection{Polarizabilities at $\mathbf{589.3 \, \mathrm{\mathbf{nm}}}$}

\begin{table}[h!]
\caption{Deviations of polarizabilities at $589.3 \, \mathrm{nm}$ calculated with various methods from the CCSD results (in au) averaged over all 41 molecules.}
\label{tab:statan589}
\centering
\begin{tabular}{lrrr}
\hline
\textbf{Method} & \textbf{MD} & \textbf{MAD} & \textbf{StdDev} \\ \hline
RPA             & $-1.61$                     & $1.82$                           & $1.70$                       \\
RPA(D)          & $0.87$                      & $0.89$                           & $0.67$                      \\
HRPA            & $-8.57$                     & $8.57$                           & $5.33$                      \\
HRPA(D)         & $-0.66$                     & $0.75$                           & $0.88$                      \\
SOPPA           & $1.60$                       & $1.61$                           & $1.25$                      \\
SOPPA(CC2)      & $1.47$                      & $1.48$                           & $1.13$                      \\
SOPPA(CCSD)     & $0.63$                      & $0.72$                           & $0.88$                      \\ \hline
\end{tabular}
\end{table}

\begin{figure}[h!]
    \centering
    \includegraphics[width=0.9\linewidth]{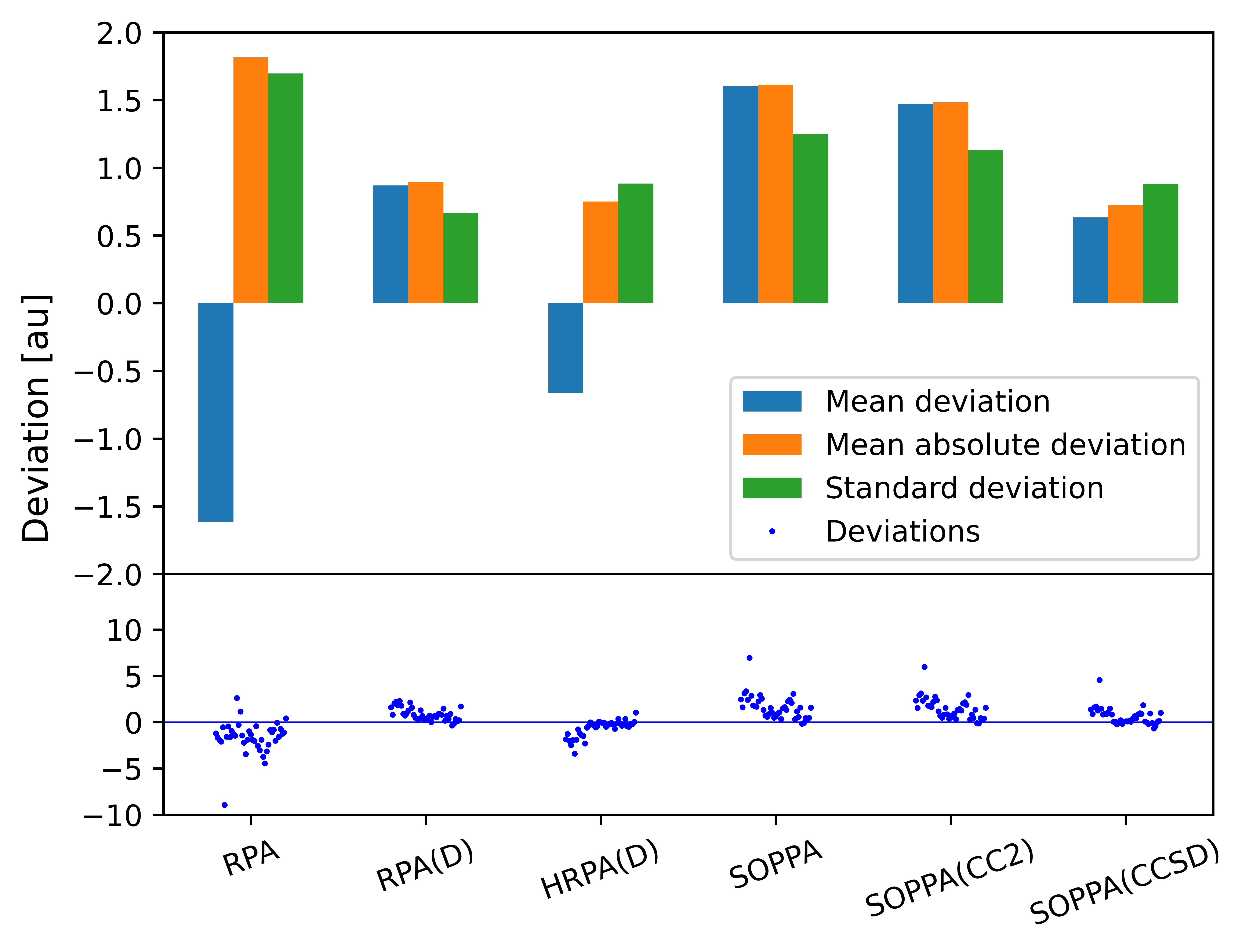}
    \caption{Deviations of the polarizabilities  at $589.3 \, \mathrm{nm}$ calculated with various methods from the CCSD results (in au) averaged over all 41 molecules.}
    \label{fig:deviation589}
\end{figure}

Next, the deviations of the calculations from the CCSD results at $589.3 \, \mathrm{nm}$ ($0.077318 \, \mathrm{au}$) are analyzed in Table \ref{tab:statan589} and Figure \ref{fig:deviation589}.
At $589.3 \, \mathrm{nm}$ both the mean-, mean absolute-, and standard deviations have increased for all methods compared to the static case (Table \ref{tab:statanstatic} and Figure \ref{fig:deviationstatic}). For HRPA(D), SOPPA, SOPPA(CC2), and SOPPA(CCSD), the increase is roughly $0.3 \, \mathrm{au}$, and for RPA(D), the increase in numerical mean- and mean absolute deviation is $0.16 \, \mathrm{au}$, and the increase in standard deviation is $0.09 \, \mathrm{au}$. HRPA has suffered an increase of $0.70 \, \mathrm{au}$ in mean- and mean absolute deviation and $0.63 \, \mathrm{au}$ in standard deviation. The increase in mean- and mean absolute deviation for RPA is quite small at $0.04 \, \mathrm{au}$ and $0.11 \, \mathrm{au}$, respectively, but it has suffered a slight increase in standard deviation of $0.24 \, \mathrm{au}$.

HRPA is inferior to all other methods. Remarkably, the numerical mean deviation for SOPPA is only $0.01 \, \mathrm{au}$ lower than that of RPA, but the mean absolute deviation is $0.21 \, \mathrm{au}$ lower for SOPPA than for RPA. This suggests that SOPPA tends to overestimate the polarizabilities, whereas RPA underestimates them; however, SOPPA is more consistent than RPA. This is confirmed by SOPPA having a lower standard deviation than RPA, and SOPPA is therefore superior to RPA. The CC2 amplitudes in SOPPA(CC2) yield results similar to, but slightly better than, the MP2 correlation coefficients in SOPPA. They are outperformed by RPA(D), but SOPPA(CCSD) and HRPA(D) are more accurate than RPA(D). However, RPA(D) is more consistent than HRPA(D) and SOPPA. SOPPA(CCSD) outperforms HRPA(D) in terms of accuracy, but they have the same consistency.

As in the static case, the aromatic molecules are responsible for the largest negative deviations when using HRPA(D). Likewise, they are responsible for most of the largest deviations when using RPA(D) - this behavior is even more pronounced here than in the static electric field. Again, no such tendency is observed with RPA. Using SOPPA, SOPPA(CC2), and SOPPA(CCSD), the deviations of the aromatic molecules are also in the upper range. The data point showing a significantly higher deviation than the remaining data points stems from cytosine. To quantify the effect on the statistical data, the aromatic molecules are removed again.

\begin{table}[h!]
\caption{Deviations of polarizabilities at $589.3 \, \mathrm{nm}$ calculated with various methods from the CCSD results (in au) averaged over the 30 non-aromatic molecules.}
\label{tab:statan589nonaromatic}
\centering
\begin{tabular}{lrrr}
\hline
\textbf{Method} & \textbf{MD} & \textbf{MAD} & \textbf{StdDev} \\ \hline
RPA         & $-1.71$            & $1.99$                  & $1.95$             \\
RPA(D)      & $0.65$             & $0.68$                  & $0.56$             \\
HRPA        & $-6.21$            & $6.21$                  & $3.79$             \\
HRPA(D)     & $-0.28$            & $0.41$                  & $0.67$             \\
SOPPA       & $1.32$             & $1.33$                  & $1.30$              \\
SOPPA(CC2)  & $1.18$             & $1.19$                  & $1.15$             \\
SOPPA(CCSD) & $0.42$             & $0.54$                  & $0.92$                      \\ \hline
\end{tabular}
\end{table}

\begin{figure}[h!]
    \centering
    \includegraphics[width=0.9\linewidth]{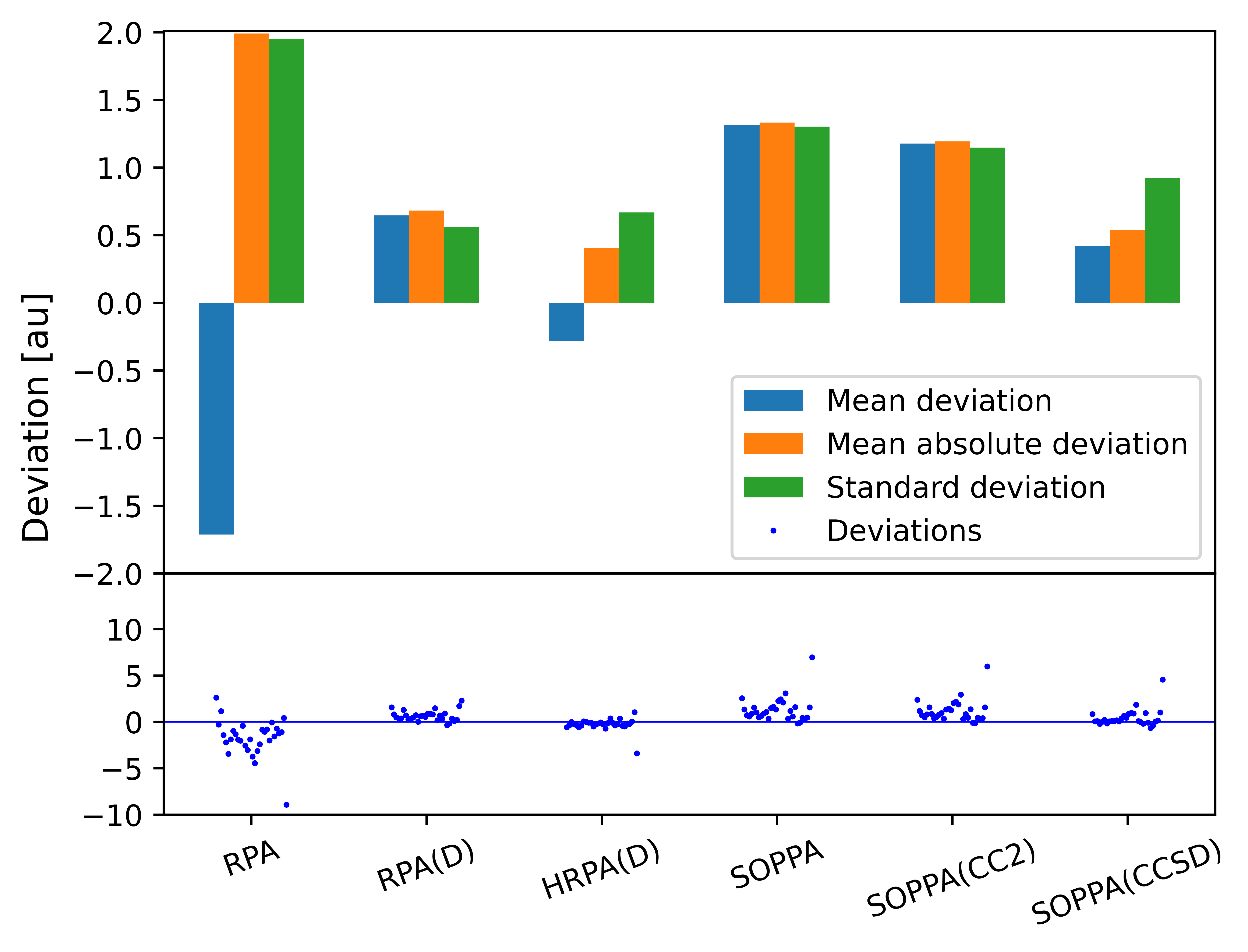}
    \caption{Deviations of the polarizabilities  at $589.3 \, \mathrm{nm}$ calculated with various methods from the CCSD results (in au) averaged over the 30 non-aromatic molecules.}
    \label{fig:deviation589nonaromatic}
\end{figure}

The statistical analysis performed on the deviations of the non-aromatic molecules (Table \ref{tab:statan589nonaromatic} and Figure \ref{fig:deviation589nonaromatic}), as expected, shows that the aromatic molecules have a significant effect at $589.3 \, \mathrm{nm}$ too. The method least affected by the aromatic molecules is RPA, whose numerical mean- and mean absolute deviation increased by only $0.17 \, \mathrm{au}$ and $0.10 \, \mathrm{au}$, respectively. However, the standard deviation has increased by $0.25 \, \mathrm{au}$. HRPA(D) is the best method, with a somewhat similar, but clearly better, performance than SOPPA(CCSD). SOPPA(CCSD) actually has a higher standard deviation than RPA(D), with a difference of $0.36 \, \mathrm{au}$. However, the mean- and mean absolute deviations for SOPPA(CCSD) are lower than those of RPA(D) by $0.23 \, \mathrm{au}$ and $0.14 \, \mathrm{au}$, respectively. RPA(D) is superior to SOPPA and SOPPA(CC2), of which SOPPA(CC2) is slightly better. Both SOPPA and SOPPA(CC2) are superior to RPA, which, in turn, is superior to HRPA. HRPA is the worst method. The order of increasing accuracy of the methods with non-aromatic molecules is thus
\begin{equation*}
    \begin{split}
        \mathrm{HRPA < RPA < SOPPA < SOPPA(CC2) <
        RPA(D) < SOPPA(CCSD) < HRPA(D)}
    \end{split}
\end{equation*}
while the order of increasing consistency is
\begin{equation*}
    \begin{split}
        \mathrm{HRPA < RPA < SOPPA < SOPPA(CC2) <
        SOPPA(CCSD) < HRPA(D) < RPA(D)}
    \end{split}
\end{equation*}

The graphical illustration of the deviations (Figure \ref{fig:deviation589nonaromatic}) reveals a data point with a much lower deviation, but higher numerical deviation, than the remaining data points for RPA, and with a significantly higher deviation for SOPPA, SOPPA(CC2), and SOPPA(CCSD). This data point, again, stems from cytosine. However, since the excitation energy for cytosine (Table \ref{SI-excbrcyt}) is substantially higher than this frequency, removing cytosine from the data set cannot be justified. The inclusion of cytosine has decreased the mean deviation, and as RPA yields polarizabilities lower than CCSD does for most molecules, it has increased the mean absolute deviation. Additionally, the standard deviation has been increased due to the inclusion of cytosine. Likewise, for SOPPA, SOPPA(CC2), and SOPPA(CCSD), the inclusion of cytosine has increased the mean-, mean absolute-, and standard deviations.

\begin{table}[h!]
\caption{Deviations of polarizabilities at $589.3 \, \mathrm{nm}$ calculated with various methods from the CCSD results (in au) averaged over the 11 aromatic molecules.}
\label{tab:statan589aromatic}
\centering
\begin{tabular}{lrrr}
\hline
\textbf{Method} & \textbf{MD} & \textbf{MAD} & \textbf{StdDev} \\ \hline
RPA         & $-1.34$            & $1.34$                  & $0.50$              \\
RPA(D)      & $1.47$             & $1.47$                  & $0.53$             \\
HRPA        & $-15.00$            & $15.00$                  & $3.19$             \\
HRPA(D)     & $-1.69$            & $1.69$                  & $0.49$             \\
SOPPA       & $2.38$             & $2.38$                  & $0.59$             \\
SOPPA(CC2)  & $2.28$             & $2.28$                  & $0.52$             \\
SOPPA(CCSD) & $1.22$             & $1.22$                  & $0.31$                      \\ \hline
\end{tabular}
\end{table}

\begin{figure}[h!]
    \centering
    \includegraphics[width=0.9\linewidth]{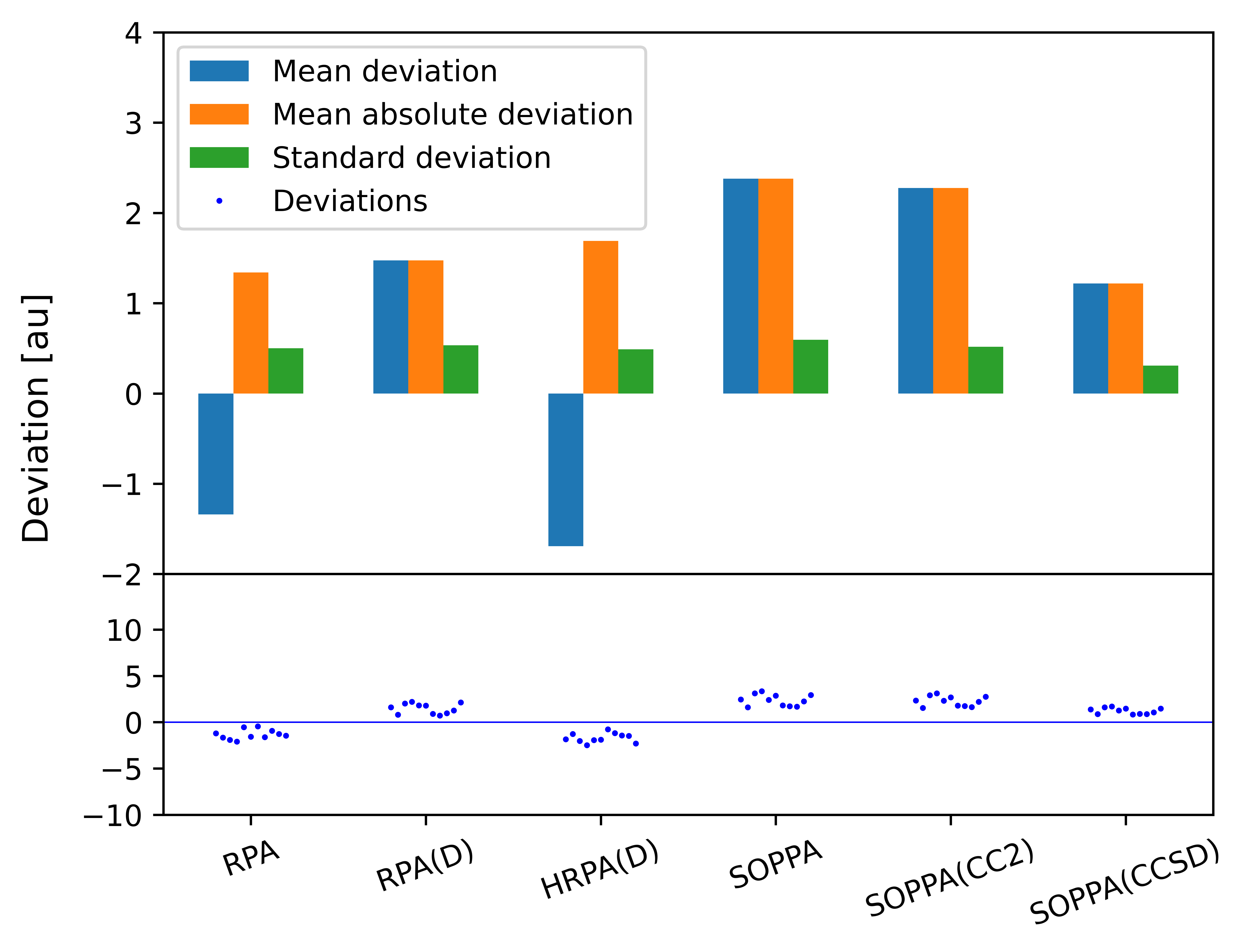}
    \caption{Deviations of the polarizabilities  at $589.3 \, \mathrm{nm}$ calculated with various methods from the CCSD results (in au) averaged over the 11 aromatic molecules.}
    \label{fig:deviation589aromatic}
\end{figure}

The performance of the methods when using only aromatic molecules is now investigated.
As expected, the deviations have become significantly worse compared to the data set with non-aromatic molecules (Table \ref{tab:statan589nonaromatic} and Figure \ref{fig:deviation589nonaromatic}). The same observation as before is made, where the numerical mean deviation and the absolute deviation are the same for a given method when considering only the aromatic molecules. It is not a surprise that HRPA yields the worst results in terms of accuracy. 
The numerical mean- and mean absolute deviations for HRPA are almost three times as large as for the data set with non-aromatic molecules and almost twice as great as for the full data set. 
The standard deviation for HRPA, however, is notably smaller than with the full data set. Using CC2 amplitudes in SOPPA(CC2) yields better results than using MP2 correlation coefficients in SOPPA. As with the non-aromatic molecules, adding the doubles correction in HRPA(D) improves drastically on the HRPA results. Here, too, HRPA(D) outperforms both SOPPA and SOPPA(CC2). HRPA(D) is marginally more consistent than RPA(D), but RPA(D) is notably more accurate. Additionally, in this case, RPA yields better results than RPA(D), but the standard deviation for RPA is $0.01 \, \mathrm{au}$ higher than for HRPA(D). RPA(D) has higher mean- and mean absolute deviations than SOPPA(CCSD) by $0.25 \, \mathrm{au}$, and a higher standard deviation by $0.22 \, \mathrm{au}$. The order of increasing accuracy is therefore
\begin{equation*}
    \begin{split}
        \mathrm{HRPA < SOPPA < SOPPA(CC2) < 
        HRPA(D) < RPA(D) < RPA < SOPPA(CCSD)}
    \end{split}
\end{equation*}
while the order of increasing consistency is
\begin{equation*}
    \begin{split}
        \mathrm{HRPA < SOPPA < RPA(D) < SOPPA(CC2) <
        RPA < HRPA(D) < SOPPA(CCSD)}
    \end{split}
\end{equation*}

\subsection{Polarizabilities at $\mathbf{355.0 \, \mathrm{\mathbf{nm}}}$}

\begin{table}[h!]
\caption{Deviations of polarizabilities at $355.0 \, \mathrm{nm}$ calculated with various methods from the CCSD results (in au) averaged over 38 molecules, without bromine, chlorine, and cytosine.}
\label{tab:statan355}
\centering
\begin{tabular}{lrrr}
\hline
\textbf{Method} & \textbf{MD} & \textbf{MAD} & \textbf{StdDev} \\ \hline
RPA             & $-1.52$                     & $1.89$                           & $1.72$                      \\
RPA(D)          & $1.34$                      & $1.37$                           & $1.31$                      \\
HRPA            & $-9.90$                      & $9.90$                            & $6.31$                      \\
HRPA(D)         & $-1.28$                     & $1.34$                           & $1.43$                      \\
SOPPA           & $2.25$                      & $2.26$                           & $1.65$                      \\
SOPPA(CC2)      & $2.04$                      & $2.05$                           & $1.54$                      \\
SOPPA(CCSD)     & $1.38$                      & $1.45$                           & $1.49$                      \\ \hline
\end{tabular}
\end{table}

\begin{figure}[h!]
    \centering
    \includegraphics[width=0.9\linewidth]{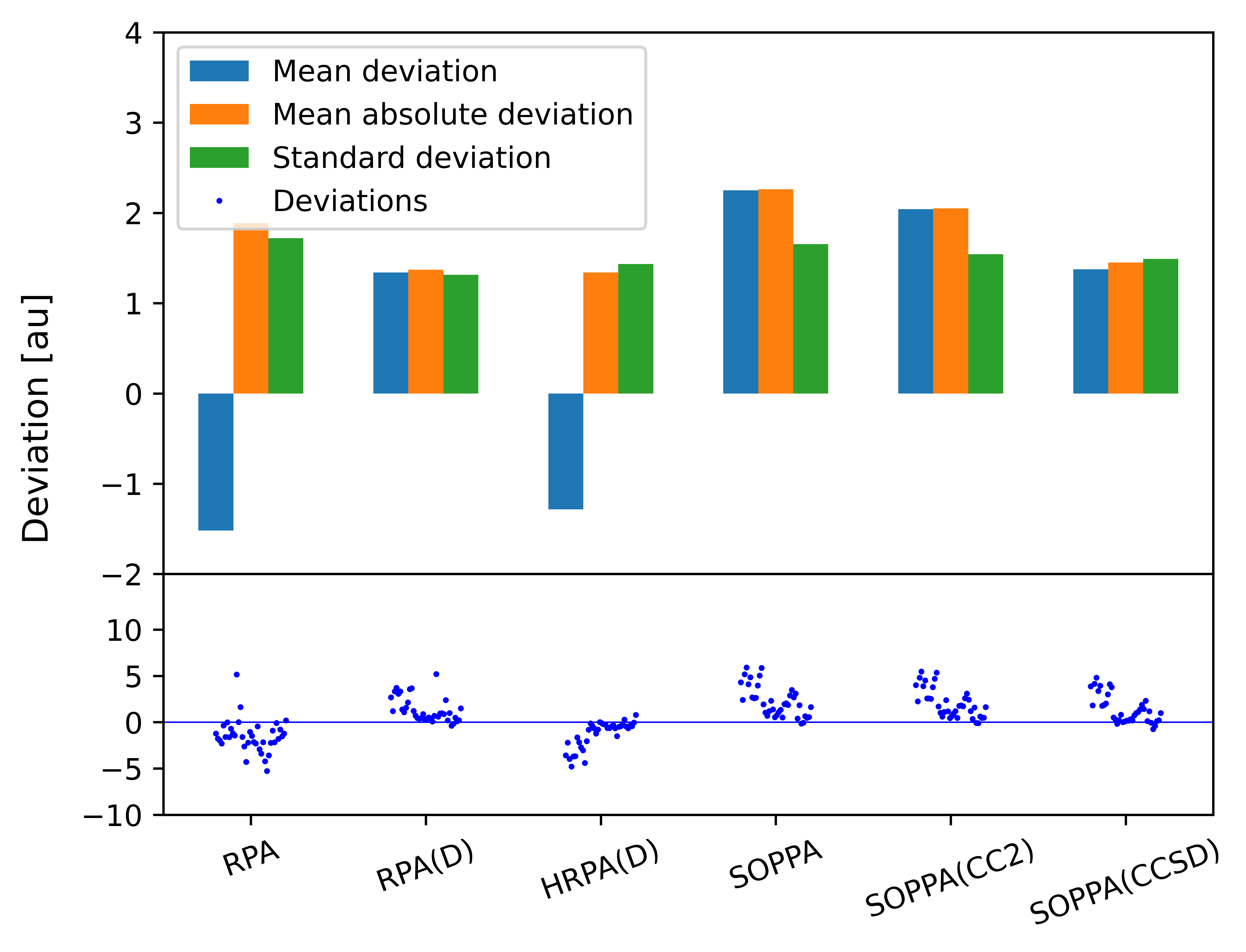}
    \caption{Deviations of the polarizabilities  at $355.0 \, \mathrm{nm}$ calculated with various methods from the CCSD results (in au) averaged over 38 molecules, without bromine, chlorine, and cytosine.}
    \label{fig:deviation355}
\end{figure}

Finally, the deviations of the calculations from CCSD results at $355.0 \, \mathrm{nm}$ ($0.128347 \, \mathrm{au}$) are analyzed in Table \ref{tab:statan355} and Figure \ref{fig:deviation355}.
Compared to the polarizabilities at $589.3 \, \mathrm{nm}$ (Table \ref{tab:statan589} and Figure \ref{fig:deviation589}), the mean-, mean absolute-, and standard deviations at $355.0 \, \mathrm{nm}$ (Table \ref{tab:statan355} and Figure \ref{fig:deviation355}) have increased for all methods, except RPA. Interestingly, for RPA, the numerical mean deviation has decreased by $0.09 \, \mathrm{au}$ and the mean absolute deviation has only increased by $0.07 \, \mathrm{au}$. 
For all other methods, the deviations have increased significantly more. 
HRPA is, as expected, the worst method by a large amount. It is outperformed by RPA, SOPPA, and SOPPA(CC2). The differences in mean- and mean absolute deviations between SOPPA and SOPPA(CC2) have increased from $589.3 \, \mathrm{nm}$, while the difference in standard deviations has actually decreased by $0.01 \, \mathrm{au}$. SOPPA(CC2) is still the better method in terms of performance. They are both slightly more consistent than RPA, but RPA is significantly more accurate. SOPPA(CCSD) has a better performance than all three methods. RPA(D) actually has lower mean-, mean absolute, and standard deviations than SOPPA(CCSD) and is, therefore, the superior method. RPA(D) is very close in performance to HRPA(D). The mean absolute deviation is greater by $0.03 \, \mathrm{au}$, however, the standard deviation is $0.12 \, \mathrm{au}$ smaller for RPA(D). HRPA is again the worst method. As with the other wavelengths, the aromatic molecules tend to have higher absolute deviations than the non-aromatic molecules. Therefore, they are removed to perform a statistical analysis on the data of only the non-aromatic molecules.

\begin{table}[h!]
\caption{Deviations of polarizabilities at $355.0 \, \mathrm{nm}$ calculated with various methods from the CCSD results (in au) averaged over 27 non-aromatic molecules, without bromine, chlorine, and cytosine.}
\label{tab:statan355nonaromatic}
\centering
\begin{tabular}{lrrr}
\hline
\textbf{Method} & \textbf{MD} & \textbf{MAD} & \textbf{StdDev} \\ \hline
RPA             & $-1.61$                     & $2.13$                           & $1.99$                      \\
RPA(D)          & $0.89$                      & $0.93$                           & $1.16$                      \\
HRPA            & $-6.49$                     & $6.49$                           & $3.04$                      \\
HRPA(D)         & $-0.48$                     & $0.56$                           & $0.53$                      \\
SOPPA           & $1.55$                      & $1.57$                           & $1.26$                      \\
SOPPA(CC2)      & $1.35$                      & $1.37$                           & $1.12$                      \\
SOPPA(CCSD)     & $0.65$                      & $0.76$                           & $0.92$                      \\ \hline
\end{tabular}
\end{table}

\begin{figure}[h!]
    \centering
    \includegraphics[width=0.9\linewidth]{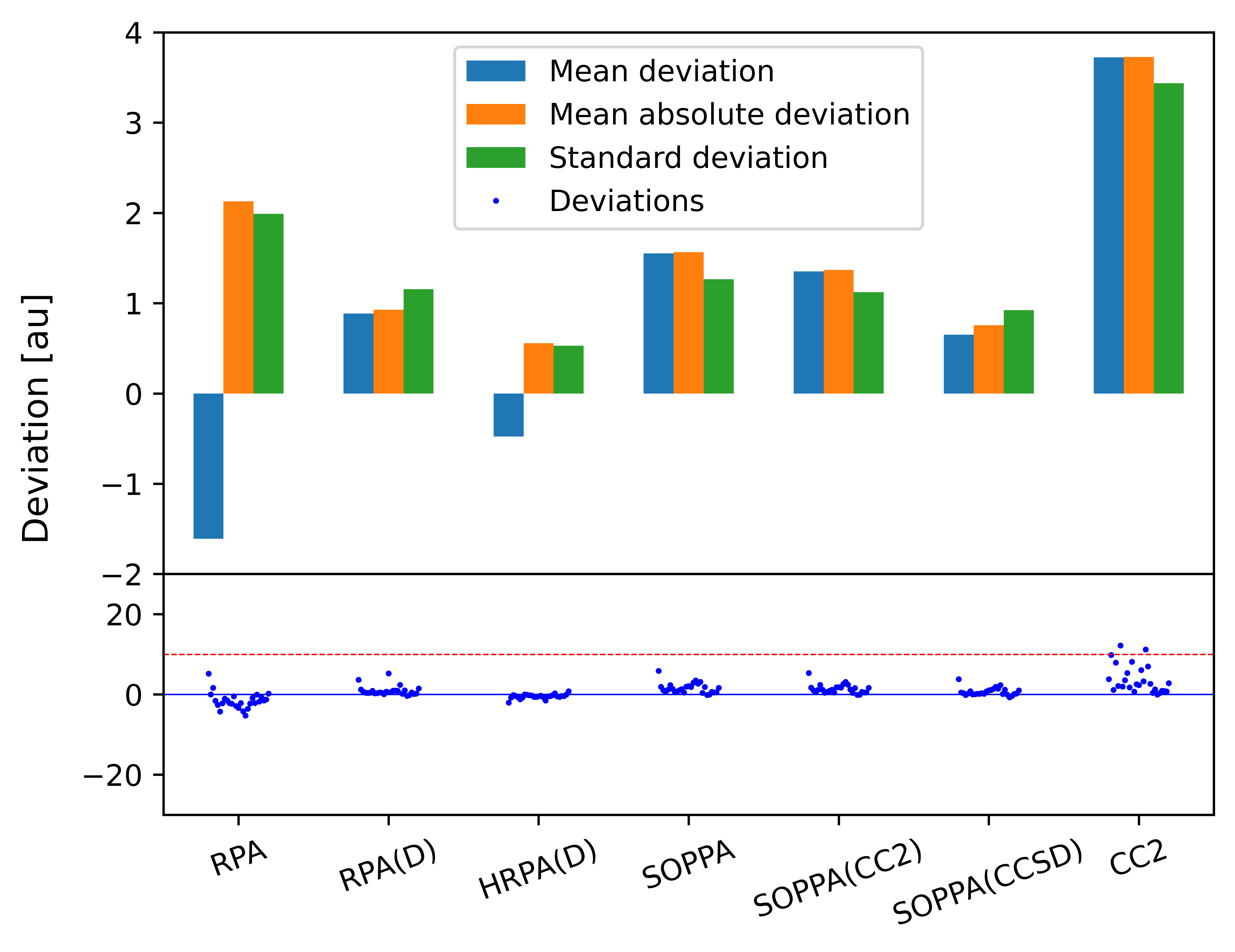}
    \caption{Deviations of the polarizabilities  at $355.0 \, \mathrm{nm}$ calculated with various methods from the CCSD results (in au) averaged over the 27 non-aromatic molecules, without bromine, chlorine, and cytosine.}
    \label{fig:deviation355nonaromatic}
\end{figure}

When considering only the non-aromatic molecules (Table \ref{tab:statan355nonaromatic} and Figure \ref{fig:deviation355nonaromatic}), all methods, except RPA, have significantly lower mean-, mean absolute-, and standard deviations. The mean-, mean absolute-, and standard deviations have, quite notably, increased for RPA. HRPA(D) is the best method for non-aromatic molecules, followed by SOPPA(CCSD). SOPPA(CC2) outperforms SOPPA, and both are better than RPA. Again, adding the doubles correction in RPA(D) improves the performance drastically, and RPA(D) is superior to SOPPA and SOPPA(CC2) in terms of accuracy. However, RPA(D) is marginally less consistent than SOPPA(CC2). RPA is considerably better than HRPA. The order of increasing accuracy is therefore
\begin{equation*}
    \begin{split}
        \mathrm{HRPA < RPA < SOPPA < SOPPA(CC2) <
        RPA(D) < SOPPA(CCSD) < HRPA(D)}
    \end{split}
\end{equation*}
while the order of increasing consistency is
\begin{equation*}
    \begin{split}
        \mathrm{HRPA < RPA < SOPPA < RPA(D) <
        SOPPA(CC2) < SOPPA(CCSD) < HRPA(D)}
    \end{split}
\end{equation*}

\begin{table}[h!]
\caption{Deviations of polarizabilities at $355.0 \, \mathrm{nm}$ calculated with various methods from the CCSD results (in au) averaged over 28 non-aromatic molecules, including cytosine, without bromine and chlorine.}
\label{tab:statan355nonaromaticwcyt}
\centering
\begin{tabular}{lrrr}
\hline
\textbf{Method} & \textbf{MD} & \textbf{MAD} & \textbf{StdDev} \\ \hline
RPA         & $-1.98$            & $2.48$                  & $2.75$             \\
RPA(D)      & $0.96$             & $1.00$                   & $1.20$              \\
HRPA        & $-7.29$            & $7.29$                  & $5.13$             \\
HRPA(D)     & $-0.74$            & $0.82$                  & $1.47$             \\
SOPPA       & $2.08$             & $2.10$                   & $3.04$             \\
SOPPA(CC2)  & $1.78$             & $1.79$                  & $2.46$             \\
SOPPA(CCSD) & $1.20$              & $1.30$                   & $3.01$                      \\ \hline
\end{tabular}
\end{table}

\begin{figure}[h!]
    \centering
    \includegraphics[width=0.9\linewidth]{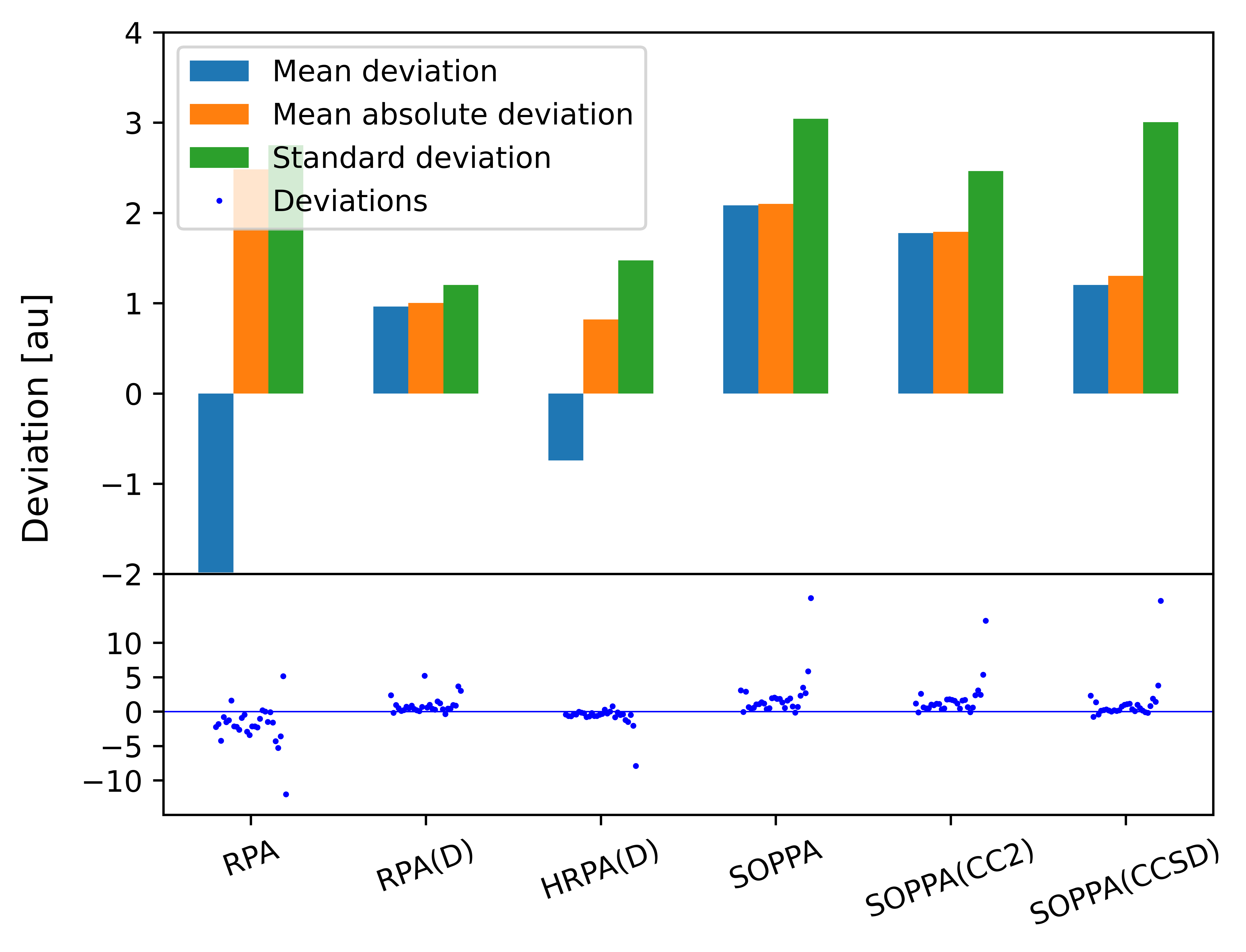}
    \caption{Deviations of the polarizabilities  at $355.0 \, \mathrm{nm}$ calculated with various methods from the CCSD results (in au) averaged over 28 non-aromatic molecules, including cytosine, without bromine and chlorine.}
    \label{fig:deviation355nonaromaticwcyt}
\end{figure}

Since the lowest excitation energy of cytosine (Table \ref{SI-excbrcyt}) is still higher than this frequency, the effects of including cytosine will also be investigated.
Compared to the data set without cytosine (Table \ref{tab:statan355nonaromatic} and Figure \ref{fig:deviation355nonaromaticwcyt}), including cytosine (Table \ref{tab:statan355nonaromaticwcyt} and Figure \ref{fig:deviation355nonaromaticwcyt}) shows an increase in numerical mean-, mean absolute-, and standard deviations for all methods. The worst method is HRPA, followed by RPA in terms of accuracy and SOPPA in terms of consistency. SOPPA(CC2) and SOPPA(CCSD) are better than SOPPA, but SOPPA(CCSD) is more accurate than SOPPA(CC2), while SOPPA(CC2) is more consistent. RPA is less consistent than SOPPA(CC2) but more consistent than SOPPA(CCSD). RPA(D) outperforms HRPA(D) in terms of consistency, but HRPA(D) is more accurate than RPA(D). Both methods outperform the remaining methods. The order of increasing accuracy is
\begin{equation*}
    \begin{split}
        \mathrm{HRPA < RPA < SOPPA < SOPPA(CC2) <
        SOPPA(CCSD) < RPA(D) < HRPA(D)}
    \end{split}
\end{equation*}
while the order of increasing consistency is
\begin{equation*}
    \begin{split}
        \mathrm{HRPA < SOPPA < SOPPA(CCSD) <
        RPA < SOPPA(CC2) < HRPA(D) < RPA(D)}
    \end{split}
\end{equation*}

\begin{table}[h!]
\caption{Deviations of polarizabilities at $355.0 \, \mathrm{nm}$ calculated with various methods from the CCSD results (in au) averaged over the 11 aromatic molecules.}
\label{tab:statan355aromatic}
\centering
\begin{tabular}{lccc}
\hline
\textbf{Method} & \textbf{MD} & \textbf{MAD} & \textbf{StdDev} \\ \hline
RPA             & $-1.29$                     & $1.29$                           & $0.66$                      \\
RPA(D)          & $2.46$                      & $2.46$                           & $0.97$                      \\
HRPA            & $-18.27$                    & $18.27$                          & $4.05$                      \\
HRPA(D)         & $-3.26$                     & $3.26$                           & $0.95$                      \\
SOPPA           & $3.97$                      & $3.97$                           & $1.17$                      \\
SOPPA(CC2)      & $3.73$                      & $3.73$                           & $1.05$                      \\
SOPPA(CCSD)     & $3.15$                      & $3.15$                           & $1.06$                      \\ \hline
\end{tabular}
\end{table}

\begin{figure}[h!]
    \centering
    \includegraphics[width=0.9\linewidth]{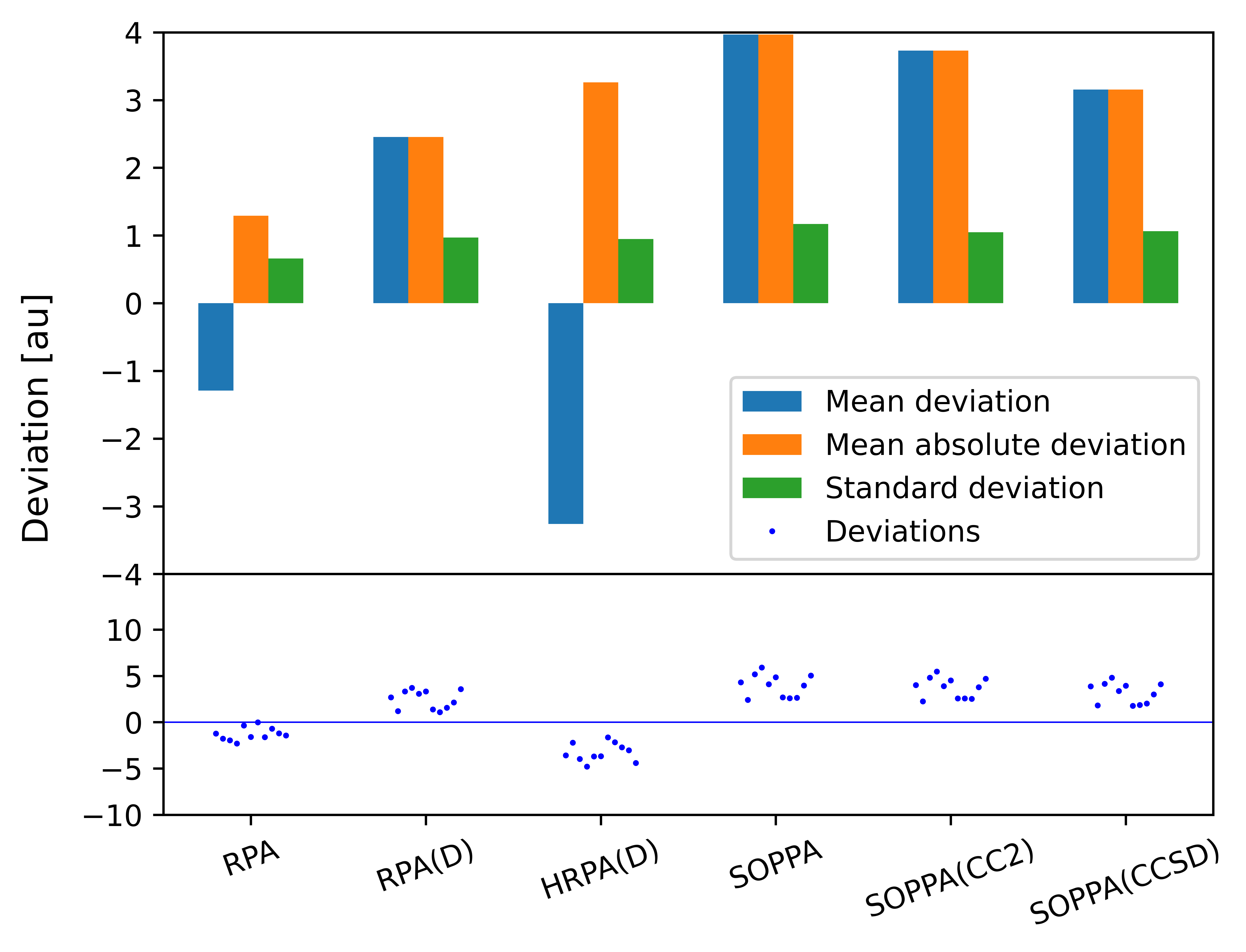}
    \caption{Deviations of the polarizabilities  at $355.0 \, \mathrm{nm}$ calculated with various methods from the CCSD results (in au) averaged over the aromatic molecules.}
    \label{fig:deviation355aromatic}
\end{figure}

Now, the data set with only the 11 aromatic molecules is considered in Table \ref{tab:statan355aromatic} and Figure \ref{fig:deviation355aromatic}.
The mean- and mean absolute deviations of all methods except RPA have increased significantly compared to the data set with non-aromatic molecules (Table \ref{tab:statan355nonaromatic} and Figure \ref{fig:deviation355nonaromatic}). For RPA, however, the numerical mean deviation has decreased by $0.32 \, \mathrm{au}$, and the mean absolute deviation has decreased by $0.84 \, \mathrm{au}$. As opposed to lower frequencies, the standard deviations have increased for only HRPA, HRPA(D), and SOPPA(CCSD), whereas it has decreased for the remaining methods. The decrease is small for all methods except RPA, for which the standard deviation is a third of what it is when considering only the non-aromatic molecules. The observation can be made here, too, that the numerical mean deviations and mean absolute deviations are the same for a given method. RPA is convincingly the best method, followed by RPA(D) in terms of accuracy and by HRPA(D) in terms of consistency. RPA(D), interestingly, is more accurate than both HRPA(D) and SOPPA(CCSD), however, only outperforms SOPPA(CCSD) in terms of consistency. The difference in consistency between these three methods is marginal, but it is more pronounced in terms of accuracy. HRPA(D) and SOPPA(CCSD) are very close in performance, with the numerical mean- and mean absolute deviations for SOPPA(CCSD) being lower than for HRPA(D) by $0.11 \, \mathrm{au}$. However, the standard deviation for HRPA(D) is $0.11 \, \mathrm{au}$ lower than for SOPPA(CCSD). Using CC2 amplitudes instead of CCSD amplitudes in SOPPA(CC2) gives higher mean- and mean absolute deviations. Interestingly, the standard deviation for SOPPA(CC2) is lower than for SOPPA(CCSD) by $0.01 \, \mathrm{au}$. Using MP2 correlation coefficients instead of Coupled-Cluster coefficients yields worse results than using CC2 amplitudes. Employing the MP2 wavefunction in HRPA, as expected, yields the worst results. The order of increasing accuracy is thus
\begin{equation*}
    \begin{split}
        \mathrm{HRPA < SOPPA < SOPPA(CC2) <
        HRPA(D) < SOPPA(CCSD) < RPA(D) < RPA}
    \end{split}
\end{equation*}
whereas the order of increasing consistency is
\begin{equation*}
    \begin{split}
        \mathrm{HRPA < SOPPA < SOPPA(CCSD) <
        SOPPA(CC2) < RPA(D) < HRPA(D) < RPA}
    \end{split}
\end{equation*}
Since the difference in consistency between HRPA(D), RPA(D), SOPPA(CCSD), and SOPPA(CC2) is so marginal, the quality of the results is likely better determined by the accuracy of the methods.

\subsection{Previous studies}

For the calculation of static polarizabilities of aromatic molecules, HRPA(D) was previously found to be more accurate and consistent than RPA, RPA(D), HRPA, and SOPPA, with CC3 results as reference data.\cite{8RPAHRPASOPPA}
Additionally, the performance of SOPPA was found to be similar to, but marginally worse than, that of SOPPA(CC2), with SOPPA(CCSD) outperforming both methods, also with CC3 results as reference data.\cite{9SOPPA}
In the same study, CCSD was superior to SOPPA(CCSD). For SOPPA, SOPPA(CC2), SOPPA(CCSD), and CCSD, the same order in performance was observed with dynamic polarizabilities, but with overall worse performances. RPA was shown to be affected less than RPA(D), HRPA, HRPA(D), and SOPPA when moving from static to dynamic polarizabilities.\cite{8RPAHRPASOPPA}
In the current study, SOPPA(CCSD) was found to have a significantly higher accuracy than RPA(D) and HRPA(D) for static polarizabilities, but it has not previously been directly compared to these two methods, as only SOPPA was included in the previously mentioned study. However, the performance of SOPPA was shown to be relatively close to that of HRPA(D),\cite{8RPAHRPASOPPA} and using CCSD amplitudes in SOPPA(CCSD) considerably improved the results compared to using MP2 correlation coefficients SOPPA.\cite{9SOPPA} 
Therefore, it can be speculated that, according to those results, SOPPA(CCSD) has a similar or better performance than HRPA(D). 
This would be in excellent correspondence with the results of the current study, with SOPPA(CCSD) convincingly being the most accurate method for static polarizabilities of aromatic molecules. 
As in previous studies, HRPA(D) was found to be more consistent than RPA, RPA(D), HRPA, and SOPPA. RPA(D), however, was previously found to have a consistency similar to that of HRPA,\cite{8RPAHRPASOPPA} which was not seen in this study. 
Instead, its consistency was found to be close to that of HRPA(D). 
The order of increasing consistency of SOPPA, SOPPA(CC2), and SOPPA(CCSD) is the same here as previously observed,\cite{9SOPPA} however, the difference between the methods is much more pronounced in the current study. 
The excellent performance of RPA that was observed with the calculation of static polarizabilities, \cite{8RPAHRPASOPPA} was not directly observed in the current study. 
However, this may be due to the inclusion of cytosine, whose numerical deviation from CCSD results was much greater than those of the remaining aromatic molecules (Figure \ref{fig:deviationstaticaromatic}). If cytosine had been removed, the results of this study for RPA might have been more in agreement with previous studies. 

Moving from static to dynamic polarizabilities was shown in this study to affect RPA less than the remaining methods, which is in excellent agreement with previous results.\cite{8RPAHRPASOPPA} RPA certainly is not the best performing method at $589.3 \, \mathrm{nm}$ in this study, however, as discussed, the inclusion of cytosine likely caused this. Had cytosine not been included, the results would be in much better agreement with previous studies. \cite{8RPAHRPASOPPA} The order in performance of SOPPA, SOPPA(CC2), and SOPPA(CCSD) is the same as previously observed with dynamic polarizabilities. \cite{9SOPPA} The difference in accuracy is similar to what was previously observed, however, the difference in consistency is more pronounced in this study. The performances of RPA(D) and HRPA(D) in this study are not fully in agreement with previous results for dynamic polarizabilities.\cite{9SOPPA} Previously, HRPA(D) was found to be slightly more accurate and notably more consistent than RPA(D). In this study, however, the difference in standard deviation is only $0.10 \, \mathrm{au}$ and $0.02 \, \mathrm{au}$ at $589.3 \, \mathrm{nm}$ and $355.0 \, \mathrm{au}$, respectively, with RPA(D) being the most consistent method. RPA(D) was also found to be more accurate than HRPA(D).

\subsection{Comparison with experimental data}

The results of the different methods are in the following compared to experimental data. In the calculations, temperature effects, vibrational contributions and possible solvent effects were not included.\cite{Russell1995, spasB12, Cammi1997}
The comparison with experimental data is therefore not without problems, but it is nonetheless included for completeness.

\begin{table}[h!]
\caption{Deviations of static polarizabilities calculated with various methods from the experimental reference data (in au).}
\label{tab:statanexp}
\centering
\begin{tabular}{lrcc}
\hline
\textbf{Method} & \textbf{MD} & \textbf{MAD} & \textbf{StdDev} \\ \hline
RPA             & $-1.86$                     & $4.94$                           & $8.13$                      \\
RPA(D)          & $0.41$                      & $4.06$                           & $8.01$                      \\
HRPA            & $-8.16$                     & $9.31$                           & $8.37$                      \\
HRPA(D)         & $-0.72$                     & $4.31$                           & $7.94$                      \\
SOPPA           & $0.97$                      & $4.26$                           & $8.28$                      \\
SOPPA(CC2)      & $0.86$                      & $4.24$                           & $8.25$                      \\
SOPPA(CCSD)     & $0.01$                      & $4.23$                           & $8.21$                      \\
CCSD            & $-0.29$                     & $4.17$                           & $8.07$                      \\ \hline
\end{tabular}
\end{table}

\begin{figure}[h!]
    \centering
    \includegraphics[width=0.9\linewidth]{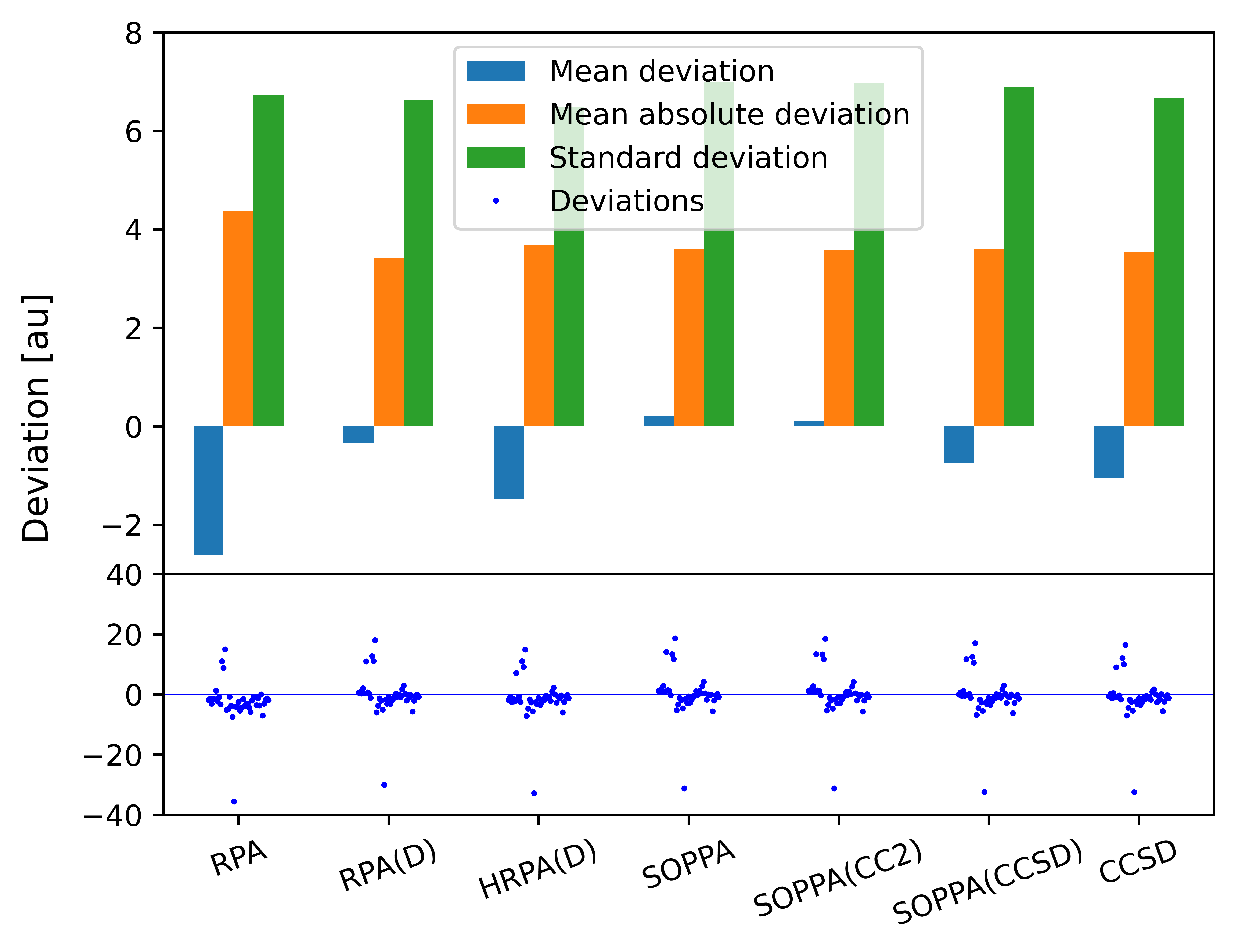}
    \caption{Deviations of the static polarizabilities (in au) calculated with various methods from the experimental data.}
    \label{fig:deviationstaticexp}
\end{figure}

When comparing the results for the static polarizabilities from all methods and all molecules with experimental data (Figure \ref{SI-fig:deviationstaticexpout} in the SI), as expected, HRPA is much worse than the remaining methods. 
Comparison with experimental data (Table \ref{tab:statanexp} and Figure \ref{fig:deviationstaticexp}) shows that SOPPA, SOPPA(CC2), and SOPPA(CCSD) yield better agreement with experiment than RPA. 
Using CC2 amplitudes in SOPPA(CC2) has a slightly better performance than using the MP2 correlation coefficients in SOPPA. 
Likewise, CCSD amplitudes in SOPPA(CCSD) yield better results than SOPPA and SOPPA(CC2). 
Adding the doubles correction to HRPA in HRPA(D) improves the HRPA results drastically and produces the most consistent results. 
Accuracy-wise, however, HRPA(D) improves on RPA, but is worse than SOPPA. Adding the doubles correction at just the RPA level instead, in RPA(D), has a better correspondence with the experimental data. 
RPA(D) gives the best agreement with the experimental values, and only HRPA(D) is more consistent with a difference in standard deviation of $0.07 \, \mathrm{au}$. CCSD gives slightly less good agreement than RPA(D), but is superior to the other methods. The standard deviations for CCSD and RPA(D) are also similar, with RPA(D) having the lowest standard deviation by $0.06 \, \mathrm{au}$. CCSD is thus outperformed in consistency by RPA(D) and HRPA(D), but is superior to the remaining methods.

The graphical representation of the deviations in Figure \ref{fig:deviationstaticexp} shows that for all methods except RPA, there are five molecules with notably higher deviations than the remaining molecules. These molecules have been identified as cytosine, pyrrole, thiophene, toluene, and propane. For RPA, this is only the case with pyrrole, thiophene, toluene, and propane, while the deviation of cytosine is much closer to those of the remaining molecules, though still notably higher. This could again point toward a tendency with the aromatic molecules. Likewise, in the lower range of deviations, all methods have a molecule with a much lower deviation than the remaining molecules. This molecule has been identified as trimethylamine.

Experimentally obtained frequency dependent polarizabilities have been found for furan and thiophene at the wavelengths $632.8 \, \mathrm{nm}$ in the vapor phase \cite{Furanexp2}, $632.8 \, \mathrm{nm}$ in cyclohexane solvent \cite{Furanexp3}, and $589.3 \, \mathrm{nm}$ in carbon tetrachloride solvent \cite{Exp589}. This data will be used to compare the calculations with experimental data for dynamic polarizabilities in Table \ref{tab:polexpfurthio}.

\begin{table}[h!]
\caption{Polarizabilities (in au) of furan and thiophene at $632.8 \, \mathrm{nm}$ and $589.3 \, \mathrm{nm}$, and deviations from experimental values.}
\label{tab:polexpfurthio}
\centering
\begin{tabular}{llcrrcccr}
\hline
\multicolumn{1}{c}{\textbf{Molecule}}
& \textbf{Method} 
& \multicolumn{3}{c}{$\mathbf{632.8 \, \mathrm{\textbf{nm}}}$} &&&
 \multicolumn{2}{c}{$\mathbf{589.3 \, \mathrm{\textbf{nm}}}$}
\\ 
\cline{3-9} 
\multicolumn{1}{c}{} & 
& \multicolumn{1}{c}{$\alpha$}
& \multicolumn{1}{c}{\textbf{Dev.}\footnote{\label{exp1} In cyclohexane solvent \cite{Furanexp3}}} 
& \multicolumn{1}{c}{\textbf{Dev.}\footnote{\label{exp2} In the vapor phase \cite{Furanexp2}}} 
&&& \multicolumn{1}{c}{$\alpha$} 
& \multicolumn{1}{c}{\textbf{Dev.}} \\
 \hline
\multirow{10}{*}{Furan}     
& Exp.                             
& \begin{tabular}[c]{@{}r@{}}$49.1 \pm 2.2$ \footref{exp1} \\ $49.1 \pm 0.5$ \footref{exp2}\end{tabular} & &  &&& $48.8$  & \\
& RPA                              & $49.1$                                                                    & $0.0$                  & $0.0$                 &&& $49.4$                           & $0.6$                                      \\
& RPA(D)                           & 50.4                                                                    & 1.3                  & 1.3                 &&& 50.7                           & 1.9                                      \\
& HRPA                             & 39.4                                                                    & $-9.7$                 & $-9.7$                &&& 39.5                           & $-9.3$                                     \\
& HRPA(D)                          & 48.9                                                                    & $-0.2$                 & $-0.2$                &&& 49.0                           & 0.2                                      \\
& SOPPA                            & 51.4                                                                    & 2.3                  & 2.3                 &&& 51.6                           & 2.8                                      \\
& SOPPA(CC2)                       & 51.3                                                                    & 2.2                  & 2.2                 &&& 51.6                           & 2.8                                      \\
& SOPPA(CCSD)                      & 50.4                                                                    & 1.3                  & 1.3                 &&& 50.7                           & 1.9                                      \\
& CCSD                             & 49.6                                                                    & 0.5                  & 0.5                 &&& 49.8                           & 1.0                                      \\ \hline
\multirow{10}{*}{Thiophene} 
& Exp.                             & \begin{tabular}[c]{@{}r@{}}$65.2 \pm 2.1$ \footref{exp1}\\ $64.9 \pm 0.6$ \footref{exp2}\end{tabular} &                      &                     &&& 60.6                           &                                          \\
& RPA                              & 64.9                                                                    & $-0.3$                 & 0.0                 &&& 65.2                           & 4.6                                      \\
& RPA(D)                           & 67.3                                                                    & 2.1                  & 2.4                 &&& 67.7                           & 7.1                                      \\
& HRPA                             & 51.7                                                                    & $-13.5$                & $-13.2$               &&& 51.8                           & $-8.8$                                     \\
& HRPA(D)                          & 64.7                                                                    & $-0.5$                 & $-0.2$                &&& 65.0                           & 4.4                                      \\
& SOPPA                            & 68.3                                                                    & 3.1                  & 3.4                 &&& 68.7                           & 8.1                                      \\
& SOPPA(CC2)                       & 68.2                                                                    & 3.0                  & 3.3                 &&& 68.7                           & 8.1                                      \\
& SOPPA(CCSD)                      & 67.1                                                                    & 1.9                  & 2.2                 &&& 67.5                           & 6.9                                      \\
& CCSD                             & 66.1                                                                    & 0.9                  & 1.2                 &&& 66.5                           & 5.9                                      \\ \hline
\end{tabular}
\end{table}

At $632.8 \, \mathrm{nm}$, RPA clearly has the best performance, which is in agreement with previous results, since both furan and thiophene are aromatic molecules. HRPA(D) performs similarly to RPA, but slightly worse. It does, however, perform better than CCSD. SOPPA(CCSD) is inferior to CCSD, and when considering furan, its performance is the same as for RPA(D). However, when considering thiophene, SOPPA(CCSD) is slightly better. SOPPA(CC2) has marginally better results than SOPPA, and both are much better than HRPA.

At $589.3 \, \mathrm{nm}$, HRPA is quite convincingly the worst method when comparing to experimental data. SOPPA and SOPPA(CC2) yield the same resultsboth for furan and thiophene, and both are inferior to RPA(D). RPA(D) yields the same result as SOPPA(CCSD) for furan, but SOPPA(CCSD) is slightly better when considering thiophene. RPA and HRPA(D) both perform better than CCSD, and HRPA(D) performs slightly better than RPA.

\begin{table}[h!]
\caption{Polarizability (in au) of $\mathrm{CO}$ at $632.8 \, \mathrm{nm}$ and $\mathrm{H_2O}$ at $514.5 \, \mathrm{nm}$.}
\label{PolCOH2O}
\begin{tabular}{lcrcr}
\hline
\multirow{2}{*}{\textbf{Method}} & \multicolumn{2}{c}{$\mathrm{\mathbf{CO}}$} & \multicolumn{2}{c}{$\mathrm{\mathbf{H_2O}}$} \\ \cline{2-5} 
& \multicolumn{1}{c}{$\alpha$}         & \textbf{Dev.}        & \multicolumn{1}{c}{$\alpha$}     & \textbf{Dev.}  \\ \hline
Exp.                     & $13.34$               &                      & $9.92 \pm 0.06$   &                \\
RPA                              & $12.64$               & $-0.70$              & $8.65$            & $-1.27$        \\
RPA(D)                           & $13.64$               & $0.30$               & $9.99$            & $0.07$         \\
HRPA                             & $11.01$               & $-2.33$              & $7.99$            & $-1.93$        \\
HRPA(D)                          & $13.38$               & $0.04$               & $9.81$            & $-0.11$        \\
SOPPA                            & $13.81$               & $0.47$               & $10.25$           & $0.33$         \\
SOPPA(CC2)                       & $13.79$               & $0.45$               & $10.20$           & $0.28$         \\
SOPPA(CCSD)                      & $13.56$               & $0.22$               & $9.94$            & $0.02$         \\
CCSD                             & $13.48$               & $0.14$               & $9.78$            & $-0.14$        \\ \hline
\end{tabular}
\end{table}

Experimental values were also found for some of the non-aromatic molecules, \textit{i.e.} $\mathrm{CO}$ and $\mathrm{H_2O}$ at $632.8 \, \mathrm{nm}$ and $514.5 \, \mathrm{nm}$, respectively,\cite{PolCOwater} as well as for $\mathrm{N_2}$ at seven different wavelengths \cite{PolN2}: $632.8 \, \mathrm{nm}$, $514.3 \, \mathrm{nm}$, $487.8 \, \mathrm{nm}$, $457.9 \, \mathrm{nm}$, $435.9 \, \mathrm{nm}$, $364.9 \, \mathrm{nm}$, and $351.0 \, \mathrm{nm}$. The calculated polarizabilities of $\mathrm{CO}$ and $\mathrm{H_2O}$ are compared to the experimental values in Table \ref{PolCOH2O}.
At $632.8 \, \mathrm{nm}$, HRPA(D) has the best correspondence with the experimental value, followed by CCSD. SOPPA(CCSD) has a better performance than SOPPA and SOPPA(CC2), and it also outperforms RPA(D). RPA(D), too, outperforms SOPPA and SOPPA(CC2), with SOPPA(CC2) being marginally superior to SOPPA. HRPA is convincingly the worst method, while RPA only outperforms HRPA. At $514.5 \, \mathrm{nm}$, SOPPA(CCSD) is the best method but only barely outperforms RPA(D). The performances of HRPA(D) and CCSD are close to that of RPA(D) and SOPPA(CCSD), but they are inferior to both methods, with HRPA(D) having the better correspondence with the experimental value. SOPPA and SOPPA(CC2) have notably worse correspondence than CCSD, with SOPPA(CC2) being marginally better. SOPPA and SOPPA(CC2) are followed by RPA and HRPA, where RPA is the better method.

\begin{table}[h!]
\caption{Polarizabilities (in au) of $\mathrm{N_2}$ at different wavelengths.}
\label{tab:N2exp}
\begin{tabular}{lrrrrrrr}
\hline
\multirow{2}{*}{\textbf{Method}} & \multicolumn{7}{c}{\textbf{Polarizability}}                                                                                               \\ \cline{2-8} 
                                 & \textbf{632.8 nm} & \textbf{514.3 nm} & \textbf{487.8 nm} & \textbf{457.9 nm} & \textbf{435.6 nm} & \textbf{363.9 nm} & \textbf{351.0 nm} \\ \hline
Exp.  & $11.92$  & $12.01$  & $12.03$  & $12.07$  & $12.10$   & $12.25$  & $12.29$  \\
RPA         & $11.91$  & $11.99$  & $12.01$  & $12.05$  & $12.08$  & $12.22$  & $12.26$  \\
RPA(D)      & $11.61$  & $11.69$  & $11.72$  & $11.75$  & $11.78$  & $11.93$  & $11.96$  \\
HRPA        & $9.75$   & $9.79$   & $9.81$   & $9.83$   & $9.84$   & $9.92$   & $9.94$   \\
HRPA(D)     & $11.54$  & $11.61$  & $11.63$  & $11.66$  & $11.69$  & $11.82$  & $11.85$  \\
SOPPA       & $11.81$  & $11.89$  & $11.92$  & $11.96$  & $11.99$  & $12.15$  & $12.19$  \\
SOPPA(CC2)  & $11.87$  & $11.95$  & $11.98$  & $12.02$  & $12.05$  & $12.21$  & $12.25$  \\
SOPPA(CCSD) & $11.88$  & $11.97$  & $12.00$   & $12.03$  & $12.07$  & $12.23$  & $12.27$  \\
CCSD        & $11.97$  & $12.05$  & $12.08$  & $12.12$  & $12.15$  & $12.30$   & $12.34$           \\ \hline
\end{tabular}
\end{table}

\begin{figure}[h!]
    \centering
    \includegraphics[width=0.9\linewidth]{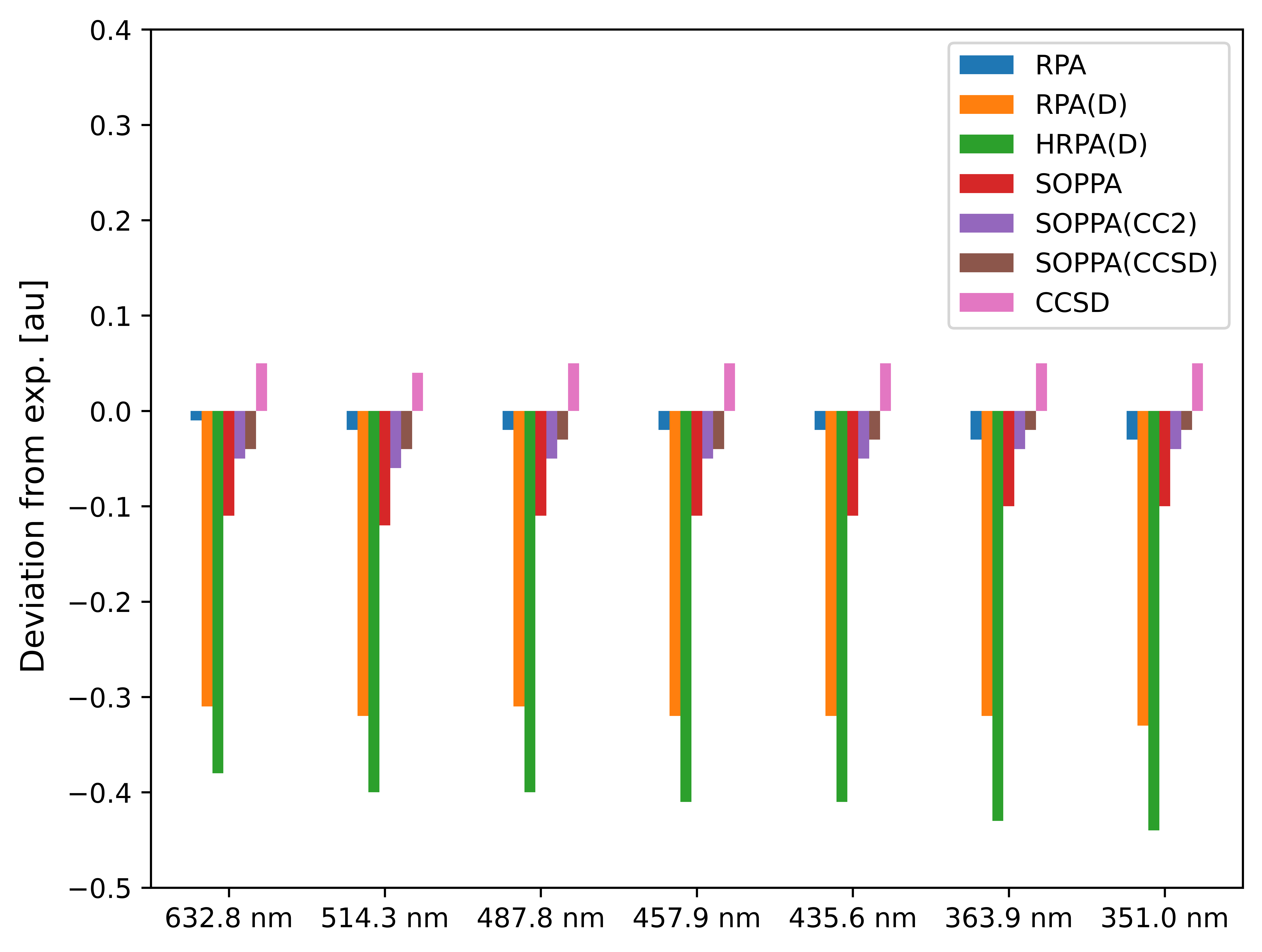}
    \caption{Deviation of calculated polarizabilities (in au) from experimental values for $\mathrm{N_2}$ at different wavelengths.}
    \label{fig:N2exp}
\end{figure}

Lastly, the calculated polarizabilities of $\mathrm{N_2}$ in Table \ref{tab:N2exp} are compared to experimental values (Figure \ref{SI-fig:N2expHRPA} in the SI). As expected, HRPA is the worst method by a large amount, so it is not shown in Figure \ref{fig:N2exp}. Apart from HRPA, HRPA(D) and RPA(D) have the worst correspondence with experimental values across all wavelengths, with HRPA(D) being the worst of the two methods. SOPPA has a much better correspondence with the experimental values, but it is also notably inferior to the remaining methods. However, it is improved on by using CC2 amplitudes in SOPPA(CC2), which subsequently is improved on by instead using CCSD amplitudes in SOPPA(CCSD). SOPPA(CCSD) seems to have a similar correspondence to the experimental values as CCSD at lower frequencies, with SOPPA(CCSD) underestimating it and CCSD overestimating it. At higher frequencies, however, SOPPA(CCSD) becomes superior to CCSD, and CCSD has a similar correspondence to the experimental values as SOPPA(CC2). At lower frequencies, RPA is better than SOPPA(CCSD), but at the two highest frequencies, SOPPA(CCSD) becomes superior to RPA.

\subsection{General performance of methods}

In the previous sections, it has been shown that there is a major difference in the performance of the chosen methods for calculating polarizabilities, when they are applied to aromatic- and non-aromatic molecules, respectively. Therefore, for a more detailed understanding, the methods will effectively be benchmarked twice with the non-aromatic molecules and the aromatic molecules, respectively. However, both data sets show a tendency for HRPA to underestimate the polarizabilities substantially. This confirms that HRPA is known to drastically overestimate the excitation energies \cite{HRPAexc} and that double excitation- and de-excitation operators are indeed necessary with an MP2 wavefunction. HRPA is the worst method for both non-aromatic and aromatic molecules, though it is much worse for aromatic molecules than for non-aromatics.

\subsubsection{Calculation of polarizabilities of non-aromatic molecules}

Using only the SCF wavefunction with single excitation- and de-excitation operators in RPA is a major improvement to HRPA. It is, however, outperformed by SOPPA, SOPPA(CC2), and SOPPA(CCSD). SOPPA and SOPPA(CC2) perform similarly, with SOPPA(CC2) being slightly better. 
Interestingly, RPA's performance becomes more similar to that of SOPPA with higher frequencies, i.e. when getting closer to the first excitation energy. 
It can therefore be hypothesized that with even higher frequencies but smaller than the first excitation energy, RPA could be superior to SOPPA and perhaps SOPPA(CC2), but this needs to be investigated further for conclusions to be drawn. 
Including the doubles correction in RPA(D) gives better results than both SOPPA and SOPPA(CC2), but is outperformed by SOPPA(CCSD). At $589.3 \, \mathrm{nm}$, RPA(D) is marginally less consistent than SOPPA(CCSD), whereas at $355.0 \, \mathrm{nm}$, it is marginally less consistent than SOPPA(CC2). At both wavelengths, there is a notable difference in consistency between SOPPA(CC2) and SOPPA(CCSD). SOPPA(CCSD) is slightly worse than adding the doubles correction to HRPA in HRPA(D). HRPA(D) consistently has the best performance of all the methods. Interestingly, with increasing frequency, the performances of all methods decrease, but this effect is much less pronounced in RPA.

\subsubsection{Calculation of polarizabilities of aromatic molecules}

In contrast to the non-aromatic molecules, applying SOPPA and SOPPA(CC2) to aromatic molecules to calculate dynamic polarizabilities yields significantly worse results than applying RPA or RPA(D). With lower frequencies, SOPPA(CCSD) is superior to RPA and has a marginally higher accuracy than RPA(D), though with a considerably lower consistency. With higher frequencies, however, the performance of SOPPA(CCSD) deteriorates. The same behavior is seen with HRPA(D) and, to a lesser degree, RPA(D). With the non-aromatic molecules, the performance of RPA decreased much more slowly than the performance of the remaining methods with increasing frequency. With the aromatic molecules, the numerical mean-, mean absolute, and standard deviations actually decrease with increasing frequency. The good performance of RPA observed here for aromatic molecules is consistent with previous findings.\cite{8RPAHRPASOPPA}

\begin{table}[h!]
\caption{Excitation energies (in au) of chlorobenzene and ammonia calculated using all methods and deviations from CCSD excitation energy.}
\label{tab:exccalc}
\centering
\begin{tabular}{lcr|cr}
\hline
\multirow{2}{*}{\textbf{Method}} & \multicolumn{2}{c|}{\textbf{Chlorobenzene}}             & \multicolumn{2}{c}{\textbf{Ammonia}}                   \\ \cline{2-5} 
& \multicolumn{1}{c}{\textbf{Exc. energy}} & \textbf{Dev.} & \multicolumn{1}{c}{\textbf{Exc. energy}} & \textbf{Dev.} \\ \hline
RPA                              & $0.210$                      & $0.024$               & $0.273$                      & $0.030$               \\
RPA(D)                           & $0.171$                      & $-0.016$              & $0.229$                      & $-0.013$              \\
HRPA                             & $0.350$                      & $0.164$               & $0.342$                      & $0.099$               \\
HRPA(D)                          & $0.156$                      & $-0.030$              & $0.222$                      & $-0.020$              \\
SOPPA                            & $0.165$                      & $-0.021$              & $0.230$                      & $-0.013$              \\
SOPPA(CC2)                       & $0.167$                      & $-0.019$              & $0.230$                      & $-0.012$              \\
SOPPA(CCSD)                      & $0.157$                      & $-0.029$              & $0.232$                      & $-0.011$              \\
CCSD                             & $0.186$                      &                        & $0.243$                      &                        \\ \hline
\end{tabular}
\end{table}

When calculating the polarizabilities of aromatic molecules, the performances of all methods except RPA drop thus rapidly with increasing frequency. With higher frequencies, RPA is therefore, with RPA(D), the best method, both in terms of performance and computational cost. 
A possible explanation for this can be found by analyzing the lowest electronic excitation energies. As observed for cytosine, a low excitation energy causes the polarizability to be much more sensitive to changes in frequency when it approaches the excitation energy. To demonstrate a difference between non-aromatic molecules and aromatic molecules, the excitation energies of chlorobenzene and ammonia, as two examples of an aromatic and non-aromatic molecule, calculated with all methods are shown in Table \ref{tab:exccalc}.
When comparing the deviations of the excitation energies for chlorobenzene and ammonia calculated with the SOPPA methods from the CCSD results, it becomes clear that only RPA and HRPA overestimate the CCSD excitation energy of both molecules, while all other methods underestimate it. 
An underestimation means that the singularity of the electronic Hessian matrix is shifted to lower frequencies compared to CCSD. When the frequency used to calculate the polarizability of the molecule approaches the frequency at which the singularity is found, the polarizability will increase strongly with the frequency and differ increasingly from the polarizability calculated with a method, whose excitation energy is at a higher frequency.
HRPA(D), SOPPA, SOPPA(CC2), and SOPPA(CCSD) predict the lowest excitation energy for chlorobenzene to be much closer to the highest frequency used in this paper than for ammonia leading to the worse performance of these methods for chlorobenzene. 
The RPA(D) excitation energy for chlorobenzene still underestimates the CCSD excitation energy but it is higher than those of HRPA(D), SOPPA, SOPPA(CC2), and SOPPA(CCSD), implying that the strong increase in the polarizability happens at higher frequencies and that the difference to the CCSD result stays smaller. 
Assuming these observations can be generalized to be true for all aromatic- and non-aromatic molecules, i.e. to molecules with and without low-lying electronic excited states, respectively, this explains the good performance of RPA and RPA(D). It also means that if the frequency increases to higher frequencies than those used in this paper, the same effect can likely be observed for non-aromatic molecules.

\section{\label{Conclusion}Conclusion}

Static- and frequency-dependent polarizabilities were calculated for a set of 41 molecules at RPA, RPA(D), HRPA, HRPA(D), SOPPA, SOPPA(CC2), and SOPPA(CCSD) levels using the aug-cc-pVTZ basis set and benchmarked against CCSD results and experimental data.

The benchmark study revealed a major difference in the performances of all methods when applying them to non-aromatic molecules and aromatic molecules, respectively. For both types of molecules and across all the applied frequencies, HRPA leads to much greater errors than the remaining methods compared to CCSD results, while HRPA(D) and SOPPA(CCSD) lead to the best results.

For static polarizabilities, results obtained with HRPA(D) for non-aromatic molecules are closest to CCSD results, followed by SOPPA(CCSD) in terms of accuracy and RPA(D) in terms of consistency. In contrast, for aromatics, SOPPA(CCSD) has the best performance. In terms of accuracy, it is followed by RPA(D) and, subsequently, HRPA(D). However, in terms of consistency, SOPPA(CCSD) is followed by HRPA(D) and SOPPA(CC2), and subsequently RPA(D) and SOPPA(CC2). 

Moving to the frequency-dependent polarizabilities, for non-aromatic molecules, HRPA(D) is still the most accurate method, followed by SOPPA(CCSD) and RPA(D), however, RPA(D) is more consistent than HRPA(D). For aromatic molecules, at lower frequencies, SOPPA(CCSD) still has the best performance. RPA is less accurate than SOPPA(CCSD) but more accurate than the remaining methods, however, HRPA(D) is more consistent than RPA. At higher frequencies, for non-aromatic molecules, HRPA(D) again is the best performing method. 
For aromatic molecules, the performance of SOPPA(CCSD) deteriorates, and RPA has the best performance, followed by RPA(D). HRPA(D) is marginally more consistent than RPA(D), but significantly less accurate.

This shows the importance of the doubles corrections in both RPA(D) and HRPA(D), allowing for results of similar or better quality than SOPPA(CCSD). Using MP2 correlation coefficients in SOPPA or CC2 amplitudes in SOPPA(CC2) is more computationally demanding. Additionally, these results have worse correspondence with CCSD results than RPA(D) and HRPA(D) do.

The good performance of RPA for aromatic molecules can be attributed to RPA being the only method, except HRPA, that overestimates the lowest electronic excitation energy.

For completeness, the calculations have also been benchmarked against experimental data. SOPPA(CCSD) was found to have the best correspondence with experimental data on static polarizabilities, and RPA and HRPA(D) were found to be the best methods for frequency-dependent polarizabilities of aromatic molecules.


\section*{Supplementary Material}
Alternative versions of the figures, where also the HRPA results are included; tables with isotropic polarizabilities calcualted with all the methods discussed; a table with the lowest excitation energies for cytosine, bromine and chlorine calculated with all the methods.

\section*{DATA AVAILABILITY}
The data that support the findings of this study are available
from the corresponding author upon reasonable request.



\providecommand{\noopsort}[1]{}\providecommand{\singleletter}[1]{#1}%

\end{document}